\documentstyle[12pt,epsf]{article}
\newcommand{\beq}{\begin{equation}}
\newcommand{\eeq}{\end{equation}}
\newcommand{\beqa}{\begin{eqnarray}}
\newcommand{\eeqa}{\end{eqnarray}}
\newcommand{\ba}{\begin{array}}
\newcommand{\ea}{\end{array}}
\newcommand{\CR}{\nonumber \\}
\newcommand{\pa}{\partial}
\newcommand{\A}{\alpha}
\newcommand{\B}{\beta}

\newcommand{\La}{\Lambda}

\newcommand{\lm}{\lambda}

\newcommand{\rep}{{\cal R}}

\newcommand{\half}{{1\over 2}}

\newcommand{\bP}{{\bf P}}
\newcommand{\bA}{{\bf A}}
\newcommand{\bB}{{\bf B}}
\newcommand{\bC}{{\bf C}}

\begin{document}

\makeatletter
\def\setcaption#1{\def\@captype{#1}}
\makeatother

\begin{titlepage}
\null
\begin{flushright} 
hep-th/9903215  \\
UTHEP-400  \\
March, 1999
\end{flushright}
\vspace{0.5cm} 
\begin{center}
{\Large \bf
N=2 Superconformal Field Theory with ADE Global Symmetry on a D3-brane Probe
\par}
\lineskip .75em
\vskip2.5cm
\normalsize
{\large Masayuki Noguchi, Seiji Terashima\footnote{Address after April 1, 
1999: Department of Physics, Faculty of Science, University of Tokyo,
Tokyo 113-0033, Japan}  and Sung-Kil Yang} 
\vskip 1.5em
{\large \it Institute of Physics, University of Tsukuba \\
Ibaraki 305-8571, Japan}
\vskip3cm
{\bf Abstract}
\end{center} \par
We study mass deformations of $N=2$ superconformal field theories with $ADE$
global symmetries on a D3-brane. The $N=2$ Seiberg-Witten curves with
$ADE$ symmetries are determined by the Type IIB 7-brane backgrounds
which are probed by a D3-brane. The Seiberg-Witten differentials $\lambda$
for these $ADE$ theories are constructed. We show that the poles of $\lambda$
with residues are located on the global sections of the bundle in an elliptic 
fibration. It is then clearly seen how the residues transform in an 
irreducible representation of the $ADE$ groups. The explicit form of $\lambda$
depends on the choice of a representation of the residues. Nevertheless 
the physics results are identical irrespective of the
representation of $\lambda$. This is considered as the global symmetry version
of the universality found in $N=2$ Yang-Mills theory with local $ADE$ 
gauge symmetries.

\end{titlepage}

\baselineskip=0.7cm

\section{Introduction}

Probing the 7-brane background of Type IIB compactification on $\bP^1$ by a 
D3-brane provides a powerful machinery to analyze the non-perturbative
behavior of four-dimensional $N=2$ supersymmetric gauge theories
\cite{Sen1,BDS}. In this setup, the space-time gauge symmetry is transmuted
into the global symmetry in the world volume $N=2$ supersymmetric gauge theory
on a D3-brane. Then it is found in \cite{Sei} that there exist 
non-trivial $N=2$ superconformal fixed points with exceptional global
symmetries. In \cite{MN1,MN2}, on the other hand, $N=2$ fixed points with $E_n$
global symmetries are considered as a natural extension of foregoing
works \cite{SW2,AD,APSW,EHIY}. 

Although the $N=2$ theory with exceptional
symmetry does not admit the Lagrangian description, recent advances in
string duality have made it possible to study the strong-coupling regime
of $N=2$ theory by the stringy technique. For instance, it requires 
a considerable amount of effort in general to analyze the 
properties of the BPS spectrum of $N=2$ theory. The junction picture of
BPS states, however, gives the simple constraint on the 
BPS spectrum \cite{MNS,dWHIZ}. With the use of this constraint, some 
characteristic features of the BPS states in $N=2$ theory with $E_n$ symmetries
are revealed \cite{dWHIZ}.

In this paper we study mass deformations of $N=2$ theories with $ADE$ global
symmetries in detail. The present work is partly motivated in our attempt to
get a clearer understanding of the results obtained by Minahan and
Nemeschansky \cite{MN1,MN2} in formulating the elliptic curves and 
the Seiberg-Witten (SW) differentials for $E_n$ theories. It was found 
in \cite{MN2} that, for a given elliptic curve, the SW differential $\lambda$
is not uniquely determined, but depends on the representations (fundamental or
adjoint) of the global symmetry group. It is then argued that $\lambda$
in different representations lead to different physics.

In our approach we proceed along the line of the D3-brane probe picture 
and discuss systematically the curves and the differentials for the $ADE$ 
theories. In particular we clarify a great deal the properties of the pole 
terms of the SW differential.
Even for the case of $N=2$ $SU(2)$ QCD with $N_f \leq 3$, which is thought
to be well understood, we gain a new insight. Consequently we are able to
show that the representations of the ADE groups from which the SW differential
is built are irrelevant to the physics results. 
In this regard, our conclusion is opposed to what is argued in \cite{MN2}. 

The paper is organized as follows.
In section 2, we see that the elliptic curves for $N=2$ $ADE$ theories on a
D3-brane are naturally identified by examining the local geometry of
singularities in the compactification of Type IIB theory on ${\bf P}^1$
with the 24 background 7-branes. 
In section 3 the BPS mass formula for $N=2$ $ADE$ theories is
discussed in the light of the string junction lattice. In section 4 the
residues of the poles of the SW differentials for our $ADE$ theories 
are shown to transform in an
irreducible representation of the global symmetry groups. This affords a 
firm foundation of somewhat empirical construction of the SW differentials
in \cite{SW2,MN1,MN2}. In section 5 the SW differentials in the fundamental
as well as the adjoint representations are obtained in the 
$A_1,\; A_2,\; D_4,\;E_6,\; E_7$ and $E_8$ theories. 
In section 6 we analyze in detail how the SW
differential behaves under the renormalization group flow from the $E_6$ 
theory to the $D_4$ theory. In section 7 it is proved that the SW periods
are independent of the representations of the global symmetry which are
chosen to construct the SW differential. 
The result in section 7 is confirmed in section 8 by further
studying $N=2$ $SU(2)$ QCD with $N_f \le 3$. Finally we conclude in
section 9.

\section{D3-brane probe and elliptic curves}

\renewcommand{\theequation}{2.\arabic{equation}}\setcounter{equation}{0}

When Type IIB theory is compactified on ${\bf P}^1$ 
with the 24 background 7-branes,
the string coupling constant $\tau =\chi +ie^{-\phi}$, where $\chi$ is a 
R-R scalar field and $\phi$ a dilaton, is determined as the modular
parameter of an elliptic curve \cite{Vafa}
\beq
y^2=x^3+f(z) x+g(z).
\label{cubic}
\eeq
Here $z$ is a complex coordinate on ${\bf P}^1$, $f$ and $g$ are polynomials 
in $z$ of degree 8 and 12, respectively. The 24 zeroes of the discriminant 
$\Delta =4f^3+27g^2$ are the transverse positions of the 24 7-branes. The
modular parameter $\tau$ is obtained from $j(\tau)=4(24f)^3/\Delta$. The 
cubic (\ref{cubic}) describes a $K3$ surface as an elliptic fibration over the
base ${\bf P}^1$.  When the positions of some 7-branes coincide the elliptic
fibration develops singularities which are well-known to follow the Kodaira 
classification \cite{Kod}. The singularity types then have a
correspondence with the $ADE$ singularities, according to which the 
$ADE$ types of gauge symmetry in Type IIB theory are identified 
\cite{Sen1,DM}. 

The connection between the $ADE$ gauge symmetry and the background 7-brane
configurations has been established by analyzing the monodromy properties
\cite{Joh,GZ}. In Type IIB theory there exist 7-branes which are mutually 
nonlocal. To distinguish them we shall refer to a 7-brane on which Type IIB 
$(p,q)$ strings can end as a $[p,q]$ 7-brane. For the purpose of describing
the $ADE$ symmetry it is sufficient to take into account
$[1,0]$, $[1,-1]$ and $[1,1]$ 7-branes which will be henceforth denoted as
$\bA$-, $\bB$- and $\bC$-branes, respectively. Let $\bA^n \bB^m \bC^\ell$
represent a set of $n$ $\bA$-, $m$ $\bB$- and $\ell$ $\bC$-branes. The $E_8$
gauge symmetry, for instance, is realized at $\tau =e^{2\pi i/3}$ when a set
of 7-branes $\bA^7\bB \bC^2$ coalesces. 
Gauge symmetries and the corresponding 7-brane configurations relevant to 
our following discussions are summarized in Table 1. We note that
$E_5=D_5$ and the brane configuration $\bA^4\bB \bC^2$ is shown to be
equivalent to $\bA^5\bB \bC$ \cite{DHIZ}.

\begin{table}
\begin{center}
\begin{tabular}{cccc} \hline
gauge    & Kodaira  & background & coupling \\
symmetry & type     & 7-branes   & constant $\tau$   \\ \hline
$E_8$    & ${\rm II}^*$  & $\bA^7\bB \bC^2$ &   $e^{2\pi i/3}    $ \\
$E_7$    & ${\rm III}^*  $ & $\bA^6\bB \bC^2$ &   $i$           \\
$E_6$    & ${\rm IV}^*$  & $\bA^5\bB \bC^2$ &   $ e^{2\pi i/3}  $ \\
$E_5$    & ${\rm I}^*_1$ & $\bA^4\bB \bC^2$ &   $ \infty        $ \\
$D_4$    & ${\rm I}^*_0$ & $\bA^4\bB \bC$   &    arbitrary        \\
$A_2$    & ${\rm IV}  $  & $\bA^3 \bC$   &   $ e^{2\pi i/3}     $ \\
$A_1$    & ${\rm III} $  & $\bA^2 \bC$   &     $i$            \\ 
$\{ 0\}$    & ${\rm II}  $  & $\bA \bC$   &   $ e^{2\pi i/3}  $ \\ \hline
\end{tabular}
\end{center}
\caption{Symmetry and background 7-branes}
\label{tbl1}
\end{table}

We now introduce a D3-brane which is parallel to the background 7-branes.
This D3-brane can probe the local geometry near
the singularities which are
responsible for the gauge symmetry enhancement. On the D3-brane the low-energy
effective theory becomes four-dimensional $N=2$ supersymmetric gauge theory.
Suppose that the D3-brane probe is located near coalescing 7-branes, then
$N=2$ theory on the D3-brane is a fixed point theory since there are no
relevant mass parameters turned on. 
The gauge symmetry in the bulk turns out to be the enhanced global symmetry
of a fixed-point $N=2$ supersymmetric theory on the brane \cite{BDS}.

{}From this point of view, let us look at Table 1. First of all, the $D_4$
theory on the brane in the vicinity of 
the 7-branes $\bA^4\bB\bC$ arises in $N=2$ $SU(2)$ 
theory with $N_f=4$ fundamental quarks \cite{SW2}. Here $\bB$- and $\bC$-branes
stand for monopole and dyon singularities, and $\bA$-branes stand for the
squark singularities in the Coulomb branch. 
The $N=2$ $SU(2)$ theory with $N_f=4$ is
finite and the marginal gauge coupling constant can take any values.
Similarly, the $A_2$, $A_1$ and $\{ 0\}$ theories also arise 
in the Coulomb branch
of $N=2$ $SU(2)$ theory with $N_f=3$, 2 and 1, respectively. These are
non-trivial superconformal theories obtained by adjusting quark masses at
particular values \cite{APSW}. On the other hand, the $D_5$ theory describes 
the IR free behavior of $N=2$ $SU(2)$ theory with $N_f=5$.
The most interesting are the theories with $E_n$ $(n=6,7,8)$ global symmetries.
They are non-trivial $N=2$ superconformal field theories, but do not admit the
Lagrangian description. In view of the D3-brane probe approach, 
it is natural
to place these non-trivial fixed points with exceptional symmetry 
in the sequence of renormalization group flows
\beq
E_8 \; {\mathop{\hbox to 1.3cm{\rightarrowfill}}\limits_\bA} \; E_7 \;
{\mathop{\hbox to 1.3cm{\rightarrowfill}}\limits_\bA} \; E_6  \;
{\mathop{\hbox to 1.3cm{\rightarrowfill}}\limits_{\bA ,\, \bC}} \; D_4 \;
{\mathop{\hbox to 1.3cm{\rightarrowfill}}\limits_{\bA ,\, \bB}} \; A_2 \;
{\mathop{\hbox to 1.3cm{\rightarrowfill}}\limits_\bA} \; A_1 \;
{\mathop{\hbox to 1.3cm{\rightarrowfill}}\limits_\bA} \; \{ 0\},
\label{flow}
\eeq
where 7-branes indicated under the arrows are sent to infinity
to generate the flows.
In (\ref{flow}) only the $D_4$ theory is described as a local Lagrangian field
theory, while the others are considered to be non-local. Note that
the flows $E_6 \rightarrow D_4$ and $D_4 \rightarrow A_2$ are realized by
moving away mutually non-local 7-branes simultaneously. 

Starting with the $D_4$ theory one can also consider more familiar flows
\beq
D_4 \; {\mathop{\hbox to 1.3cm{\rightarrowfill}}\limits_\bA} \; D_3 \;
{\mathop{\hbox to 1.3cm{\rightarrowfill}}\limits_\bA} \; D_2  \;
{\mathop{\hbox to 1.3cm{\rightarrowfill}}\limits_\bA} \; D_1 \;
{\mathop{\hbox to 1.3cm{\rightarrowfill}}\limits_\bA} \; D_0 \, ,
\label{flowD}
\eeq
where the 7-brane background for the $D_n$ symmetry is given by
$\bA^n \bB \bC$. Note that, for $n\leq 3$, the configuration $\bA^n \bB \bC$ 
does not fall into the Kodaira classification since it is 
non-collapsible \cite{DHIZ}. On a D3-brane probing $\bA^n \bB \bC$ with
$n\leq 3$, ordinary $N=2$ $SU(2)$ QCD with $N_f=n$ fundamental quarks 
is realized.

As mentioned previously,
enhanced global symmetries at the fixed points in (\ref{flow}) are
recognized in geometric terms as the $ADE$ singularities.
Thus relevant perturbations taking the system away from criticality are
described in terms of versal deformations of the $ADE$ singularities.
The coupling constant $\tau$ of deformed $N=2$ theories is then determined by 
elliptic curves in the form of (\ref{cubic}) where the explicit forms of
polynomials $f$ and $g$ are now specified by the $ADE$ singularity types.
We have
\beqa
 E_8:& f=w_2z^3+w_8z^2+w_{14}z+w_{20},
& g=z^5+w_{12}z^3+w_{18}z^2+w_{24}z+w_{30},   \label{e8curve} \\
 E_7:&  f=z^3+w_8z+w_{12},   
&  g=w_2z^4+w_6z^3+w_{10}z^2+w_{14}z+w_{18},  \label{e7curve} \\
 E_6:&  f=w_2z^2+w_5z+w_8, 
& g=z^4+w_6z^2+w_9z+w_{12},             \label{e6curve}  \\
 D_4:&  f=z^2+\widetilde w_4, & g=w_2z^2+w_4z+w_6,     \label{d4curve} \\
 A_2:&  f=w_2,& g=z^2+w_3,   \label{a2curve}  \\
 A_1:&  f=z, & g=w_2,   \label{a1curve}
\eeqa
where the $w_q$ are deformation parameters. Here $z$ is understood as the gauge
invariant expectation value which parametrizes the vacuum moduli of $N=2$
theory. In the brane picture $z$ is a coordinate of the position of the
D3-brane probe on ${\bf P}^1$. In the cubic (\ref{cubic}) with
(\ref{e8curve})-(\ref{a1curve}) we take $y^2$ to be of degree $h$
with $h$ being the Coxeter number of $G=ADE$ (see Table 2). 
Then $x,y,z$ have the degree $q_x, q_y, q_z$ as given in Table 2 
and $w_{q_i}$ has the degree $q_i=e_i+1$ where $e_i$ is the $i$-th 
exponent of $G$. Note here that $q_x+q_z=q_y+1$ and $2q_y=h$.
The value of $q_z$ gives the scaling dimension of the expectation 
value $z$ \cite{APSW,MN1}.

\begin{table}
\begin{center}
\begin{tabular}{ccccccc} \hline
      & $E_8$ & $E_7$ & $E_6$ & $D_4$ & $A_2$ & $A_1$ \\ \hline
$h$   & 30    &  18   & 12    & 6     & 3     & 2     \\
$q_y$ & 15    &  9    & 6     & 3     & 3/2   & 1     \\
$q_x$ & 10    &  6    & 4     & 2     & 1     & 2/3   \\
$q_z$ & 6     &  4    & 3     & 2     & 3/2   & 4/3   \\ \hline
\end{tabular}
\end{center}
\caption{Degree of variables}
\label{tbl2}
\end{table}

Notice that only in the $D_4$ theory the coupling constant $\tau$ is marginal,
and hence the curve may incorporate the $\tau$-dependence. This is allowed
since $x$ and $z$ have the same degree $q_x=q_z=2$ which holds 
only for the $D_4$ case. In fact the
Seiberg-Witten (SW) curve for the $D_4$ theory obtained originally
in \cite{SW2} depends
on both $\tau$ and four bare quark masses $m_1, m_2, m_3, m_4$. 
It is not difficult to work out how the
SW curve in \cite{SW2} is related to our $D_4$ curve (\ref{d4curve}). Let us
write down the SW curve presented in (17.58), 
section 17 of \cite{SW2}\footnote{In writing (\ref{swd4}) we have replaced
$m_a$ by $m_a/2$ in (17.58) of \cite{SW2}. 
This is necessary to agree with section 16 of
\cite{SW2}. See section 17.4 of \cite{SW2}.}
\beq
Y^2=X(X-\alpha Z)(X-\beta Z)+aX^2+bX+cZ+d,
\label{swd4}
\eeq
where we have used $Z$ instead of $u$ to denote the adjoint Higgs expectation
value and
\beqa
&& a=(\alpha -\beta)^2 u_2/4, \hskip10mm
b=-(\alpha -\beta)^2 \alpha\beta u_4/4
        +i\alpha\beta (\alpha^2-\beta^2) \widetilde u_4/4,  \CR
&& c=-i(\alpha -\beta) \alpha^2\beta^2 \widetilde u_4/2,  \hskip10mm
 d=(\alpha -\beta)^2\alpha^2\beta^2u_6/4, \CR
&& \alpha =-\vartheta_3^4(\tau),  \hskip10mm \beta =-\vartheta_4^4(\tau), \CR
&& \vartheta_3(\tau)=\sum_{n \in {\bf Z}} q^{n^2/2},
\hskip10mm   \vartheta_4(\tau)=\sum_{n \in {\bf Z}} (-1)^n q^{n^2/2},
\hskip10mm q=e^{2\pi i\tau}.
\label{d4param}
\eeqa
Here the $D_4$ invariants made of quark masses are defined by
\beqa
&& u_2=-\sum_a m_a^2, \hskip10mm   u_4=\sum_{a<b}m_a^2m_b^2,  \CR
&& u_6=-\sum_{a<b<c}m_a^2m_b^2m_c^2, 
\hskip10mm  \widetilde u_4=-2im_1m_2m_3m_4.
\eeqa
Making a change of variables 
\beq
X=-\alpha\beta x, \hskip10mm
Y=\alpha\beta (\alpha -\beta)y/2,  \hskip10mm
Z=i (\alpha -\beta)z/2-(\alpha +\beta) x/2,
\label{xtoX}
\eeq
we see that (\ref{swd4}) becomes
\beq
y^2=xz^2+x^3+u_2x^2+u_4x+\widetilde u_4 z+u_6
\eeq
which is nothing but the standard form of deformations of the 
$D_4$ singularity. We next replace $x$ by $x-u_2/3$ and shuffle the
$D_4$ invariants as
\beqa
&& u_2=-3w_2, \hskip10mm   u_4=\widetilde w_4+3w_2^2, \CR
&& u_6=w_6-w_2\widetilde w_4-w_2^3, \hskip10mm   \widetilde u_4 =w_4.
\label{shuffle}
\eeqa
Then we obtain the $D_4$ curve with (\ref{d4curve}).

\section{BPS mass formula}

\renewcommand{\theequation}{3.\arabic{equation}}\setcounter{equation}{0}

Having obtained the SW curve for $N=2$ theory on a D3-brane probe, we next 
discuss the BPS mass formula. In the brane probe approach, BPS states on
the D3-brane world volume are geometrically realized as Type IIB strings, 
or more generally string junctions obeying the BPS condition. According to
\cite{dWZ}, junctions are specified by asymptotic charges $(p,q)$ and a
weight vector of $G=ADE$. Denoting a junction as ${\bf J}$ we have \cite{dWZ}
\beq
{\bf J}=p \mbox{\boldmath $\bf \omega$}^p+q \mbox{\boldmath $\bf \omega$}^q
+\sum_{i=1}^{{\rm rank}\; G} a_i \mbox{\boldmath $\bf \omega$}_i,
\label{junction}
\eeq
where $\mbox{\boldmath $\bf \omega$}^p$ and $\mbox{\boldmath $\bf \omega$}^q$ 
are junctions which are singlets
under $G$ with asymptotic charges $(1,0)$ and $(0,1)$ respectively, and the 
$\mbox{\boldmath $\bf \omega$}_i$ with zero asymptotic charges are 
junctions corresponding to
the fundamental weights of $G$. Here the $a_i$ are the Dynkin labels 
representing a weight vector. The BPS condition on ${\bf J}$ is 
described as \cite{MNS,dWHIZ}
\beq
({\bf J}.{\bf J})-{\cal GCD}(p,q) \geq -2,
\label{selection}
\eeq
where $(\; .\; )$ stands for the bilinear form on the junction 
lattice \cite{dWZ}. 

The BPS junction with $(p,q)$ charges can end on the D3-brane and realizes
the BPS state with electric $p$ and magnetic $q$ charges in the world volume
$N=2$ theory. Sen has first figured out this and, furthermore, shown how the
SW BPS mass formula in the $D_4$ theory is obtained from the mass formula
for a $(p,q)$ string in Type IIB theory \cite{Sen2}. 
His proof is easily extended to the general $ADE$ case. For this, 
let us recapitulate the basic elements in the SW theory \cite{SW1,SW2}. 
The SW differential $\lambda$ associated with an elliptic curve has to obey
\beq
{\pa \lambda \over \pa z}=\kappa {dx \over y}+d(*)
\label{cond}
\eeq
with a normalization constant $\kappa$. The SW periods are then given by
\beq
a(z)=\oint_\alpha \lambda, \hskip10mm a_D(z)=\oint_\beta \lambda,
\label{swshuuki}
\eeq
where $\alpha$ and $\beta$ are two homology cycles on a torus. The $N=2$
central charge for a BPS state with charges $(p,q)$ reads
\beq
Z= pa(z)+qa_D(z)+ {1 \over \sqrt{2}} \sum_a s_a m_a ,
\label{central}
\eeq
where the $m_a$ are the bare mass parameters and 
the $s_a$ are the global abelian charges. The BPS mass is then given by
\beq
m =\sqrt{2} |Z|.
\eeq

Let us now recall the standard elliptic function formula for the discriminant
of the cubic
\beq
\Delta (z)=-2^{20} \left( {\pi \over 2\omega_1}\right)^{12} \eta(\tau)^{24},
\label{discrim}
\eeq
where $\eta(\tau)$ is the Dedekind eta function and $2\omega_1$ is the
period along the $\alpha$-cycle
\beq
2\omega_1 = \oint_\alpha {dx \over y}.
\eeq
We thus verify the crucial formula from (\ref{discrim}) that
\beq
da(z)=\kappa \pi (-1)^{1 \over 12}2^{5 \over 3}
\eta(\tau)^2\Delta (z)^{-{1\over 12}}dz .
\label{crucial}
\eeq

In Type IIB theory on ${\bf P}^1$, on the other hand, the mass of 
a $(p,q)$ string stretched along a path $C$ is given by
\beq
m_{p,q}=\int_C T_{p,q}ds,
\eeq
where the tension of a $(p,q)$ string reads
\beq
T_{p,q}={1\over \sqrt{{\rm Im}\; \tau}} |p+q\tau |
\eeq
and the line element is given in terms of the metric
\beq
ds^2={\rm Im}\; \tau \left| \eta(\tau)^2 
\Delta (z)^{-{1\over 12}} dz \right|^2.
\eeq
A BPS state with a mass $m_{p,q}^{\rm BPS}$ is obtained by choosing a 
curve $C$ so that $C$ is a geodesic. Then, following \cite{Sen2}, one can show
$m_{p,q}^{\rm BPS} \propto m$ with the aid of (\ref{crucial}).

The BPS junctions are lifted to holomorphic curves in F/M theory compactified
on an elliptically fibered $K3$ surface. From this viewpoint, it is interesting
to see that the expression (\ref{junction}) of a junction looks quite 
similar in form to the central charge (\ref{central}). 
We may think of the $\alpha$ and $\beta$
cycles as the projection of the 
$\mbox{\boldmath $\bf \omega$}^p$ and $\mbox{\boldmath $\bf \omega$}^q$
junctions on the $x$-plane. It is obscure, however, how to understand
the bare mass term in (\ref{central}) in the light of the third term of
(\ref{junction}) which consists of the junctions associated to the
fundamental weights. In fact there is an important subtlety here. 
In massive theory, the global
abelian charges $s_a$ in (\ref{central}) carry only ``constant parts'' of
the physical abelian charges \cite{Ferr}. The periods $a,\, a_D$ can also
produce terms of constants multiple of bare masses 
\cite{Ferr,BrSt}. These terms can arise in the period integrals in massive
theory since the SW differential has the poles with residues proportional
to bare masses \cite{BF}. 
In other words, to determine the abelian charges appearing explicitly
in the $N=2$ central charge, one has to analyze the meromorphic properties of
the SW differential carefully.

\section{Residues of the Seiberg-Witten differential}

\renewcommand{\theequation}{4.\arabic{equation}}\setcounter{equation}{0}

In this section our purpose is to discuss some general properties of the
SW differential $\lambda$ associated to our $ADE$ elliptic curves with
(\ref{e8curve})-(\ref{a1curve}) for the mass deformed $ADE$ theories. 
The differential $\lambda$ satisfies
(\ref{cond}) where a normalization constant $\kappa$ will be fixed later on. 
In order to find $\lambda$ we first follow section 17.1 of \cite{SW2}. 
Let $X$ be a complex surface defined by $y^2=W(x,z; w_i)$ 
as in (\ref{e8curve})-(\ref{a1curve}). A holomorphic two-form $\Omega$ 
on the surface reads
\beq
\Omega =\kappa {dx \wedge dz \over y}.
\label{2form}
\eeq
We wish to rewrite the condition (\ref{cond}) in terms of $\Omega$. To do so,
note that, for $\lambda =a(x,z)dx$, (\ref{cond}) is written as
\beq
\kappa {dx \over y} ={\pa a(x,z) \over \pa z}dx+{\pa F(x,z) \over \pa x}dx,
\eeq
where $F(x,z)$ has appeared from the total derivative term in (\ref{cond}).
Define a one-form $\widetilde \lambda =-a(x,z)dx+F(x,z)dz$, then (\ref{cond})
is succinctly written as
\beq
\Omega =d \widetilde \lambda .
\label{close}
\eeq
This means that there exists a smooth differential $\widetilde \lambda$ obeying
(\ref{close}) if and only if the cohomology $H^2(X,{\bf C})$ is trivial.

Suppose now that $H^2(X,{\bf C})$ is non-trivial, and let the $[C_a]$ linearly
span $H^2(X,{\bf C})$. The Poincar\'e dual of $[C_a]$, which is a 
complex curve, is a non-trivial homology cycle in $X$. In this case, 
the relation (\ref{close}) is modified to be
\beq
\Omega =d \widetilde \lambda 
-2\pi i \sum_a {\rm Res}_{C_a}(\widetilde\lambda)\cdot [C_a].
\label{modclose}
\eeq
This describes the situation in which $\widetilde \lambda$ has poles on 
the $C_a$ with residues ${\rm Res}_{C_a}(\widetilde \lambda)$ and $[C_a]$ is 
a delta function supported on $C_a$. 

There is an important relation between the period integrals of $\Omega$ and
the residues \cite{SW2}. We may evaluate the periods
\beq
\pi_a =\oint_{C_a} \Omega
\label{period}
\eeq
upon compactifying $X$ in an appropriate way. Then the cohomology class
$[\Omega ]$ is expanded in terms of $[C_a]$ as
\beq
[\Omega ]= \sum_{a,b}\pi_a (M^{-1})_{ab} [C_b],
\label{expansion}
\eeq
where $M_{ab}=\sharp (C_a\cdot C_b)$ is the intersection matrix which is 
invertible. Expressing (\ref{modclose}) in cohomology and comparing to 
(\ref{expansion}) one obtains \cite{SW2}
\beq
{\rm Res}_{C_a}(\widetilde\lambda)=-{1 \over 2\pi i}\sum_b (M^{-1})_{ab} \pi_b.
\label{residue}
\eeq

Let us further examine the periods $\pi_a$. Since the defining equation for
$X$ is $y^2=W(x,z; w_i)$, the period integral (\ref{period}) takes the form
\beq
\pi_a =\kappa \oint_{C_a} {dx \wedge dz \over W(x,z; w_i)^{1/2}}.
\label{lgperiod}
\eeq
We recall here that in the Landau-Ginzburg description of two-dimensional
$ADE$ $N=2$ superconformal field theories, 
$W(x,z; w_i)$ is identified with the 
superpotential \cite{MVW}. Being twisted, these theories turn out to be
topological ones which can couple to topological gravity. Then, exactly the
same periods as (\ref{lgperiod}) have appeared when we calculate the 
one-point functions in two-dimensional gravity \cite{EYY}.
It is shown there that the periods $\pi_a$ obey the Gauss-Manin
differential equation
\beq
\left( {\pa^{2} \over \pa t_{i}\pa t_{j}}
-\sum_{k=1}^r {C_{i j}}^{k}(t) {\pa^{2} \over \pa t_{k}\pa t_{r}} \right)
\pi_a(t)=0,
\label{GM}
\eeq
where $r={\rm rank}\, G$ $(G=ADE)$, 
$t_i$ $(i=1,\cdots, r)$ are the flat coordinates
judiciously made of the $w_i$ and ${C_{i j}}^{k}(t)$ are 
the three-point functions in the $ADE$ topological Landau-Ginzburg models.
It is then clear from (\ref{residue}) that 
${\rm Res}_{C_a}(\widetilde \lambda)$ satisfy (\ref{GM}).

To find a class of solutions of the Gauss-Manin system (\ref{GM}),
we introduce
\beq
P_G^{\cal R}(t,u_i)={\rm det} (t-\Phi_{\cal R}).
\label{charapol}
\eeq
This is the characteristic polynomial in $t$ of degree ${\rm dim}\, {\cal R}$
where ${\cal R}$ is an irreducible representation of $G$.
Here $\Phi_{\cal R}$ is a representation
matrix of ${\cal R}$ and $u_i$ $(i=1,\cdots ,r)$ is the Casimir built out
of $\Phi_{\cal R}$ whose degree equals $e_i+1$ with $e_i$ being the $i$-th
exponent of $G$. (\ref{charapol}) may be solved formally with respect to the
top Casimir $u_r$, yielding
\beq
u_r=\widetilde W_G^\rep (t,u_1,\cdots ,u_{r-1}).
\label{solve}
\eeq
If we define \beq
W_G^\rep (t,u_1,\cdots, u_r)=\widetilde W_G^\rep (t,u_1,\cdots, u_{r-1})-u_r,
\label{sp}
\eeq
then $W_G^\rep (t,u_i)$ is the single-variable version of the Landau-Ginzburg
superpotential which gives rise to the same topological field theory results
with the standard $ADE$ topological Landau-Ginzburg models equipped with
the superpotential $W(x,z;w_i)$ independently of the representations 
$\rep$ \cite{EY,ItYa}. Upon doing these computations one figures out how
the Casimirs $u_i$ are related with the deformation parameters $w_i$,
and hence with the flat coordinates $t_i$.

Let $m_a$ $(a=1,\cdots ,{\rm dim}\, \rep)$ be an eigenvalue of $\Phi_\rep$,
then (\ref{charapol}) is written as
\beq
P_G^{\cal R}(t,u_1,\cdots ,u_r)=\prod_{a=1}^{{\rm dim}\, \rep}(t-m_a)
\label{factor}
\eeq
with
\beq
m_a=(\lambda_a, \phi),
\label{zero}
\eeq
where the $\lambda_a$ are the weights of $\rep$ and $(\; ,\; )$ stands for the
inner product. Here
\beq
\phi =\sum_{i=1}^r \phi_i \alpha_i
\label{vev}
\eeq
with $\alpha_i$ being the simple roots of $G$. Expanding the RHS's of
(\ref{charapol}) and (\ref{factor}) we see how the Casimirs $u_i$ are 
expressed in terms of $\phi_i$. 

In \cite{IY2}, using the
technique of topological Landau-Ginzburg models, it is shown that 
the zeroes $m_a$ of the characteristic polynomial for
any irreducible representation of the $ADE$ groups satisfy the Gauss-Manin
system (\ref{GM}) for the $ADE$ singularity. Therefore we are led to take
\beq
{\rm Res}_{C_a}(\widetilde \lambda)= \gamma_\rep \;  m_a(w), 
\hskip10mm  a=1, \cdots , {\rm dim}\,\rep ,
\label{resres}
\eeq
where $\gamma_\rep$ is a normalization constant which may depend on $\rep$.
The residues of the SW differential thus transform in the representation
$\rep$ of the global symmetry $G$.

Having fixed the residues we now would like to determine two-cycles 
$C_a$ on which the poles are located. This is the issue to which 
we turn in the next section.

\section{Seiberg-Witten differential}

\renewcommand{\theequation}{5.\arabic{equation}}\setcounter{equation}{0}

In \cite{MN1,MN2} the SW differentials in the cases of $D_4$, $E_6$, $E_7$ and
$E_8$ have been constructed by exploiting the idea of \cite{SW2} that
$y^2$ in the cubic becomes a perfect square when $x$ is at the position of
the pole. It was then found that one can obtain the SW differentials for the
adjoint in addition to the fundamental of the global symmetry group.
We wish to demonstrate that the procedure can be formulated in a more
transparent and systematic way. For this purpose, it will be shown 
in this section that the complex curves $C_a$ on which the SW
differential has poles are given by the global sections of the bundle
in an elliptic fibration, and furthermore $C_a$ have one-to-one
correspondence with the irreducible representations of the global symmetry
group $G=ADE$. The relations among the global sections in the elliptic
fibration, characteristic polynomials and algebraic equations have been
studied by Shioda in his works on the theory of Mordell-Weil 
lattice \cite{Shio}.\footnote{One of us (SKY) is indebted to K. Oguiso
for informing of Shioda's works.}

Let $(x_a(z),y_a(z))$ be such sections, then poles are located at $x=x_a(z)$
on the $x$-plane. The residues of the poles are given by (\ref{resres}) where
$m_a$ are the eigenvalues of a representation matrix $\rep$. 
Then, following Minahan and Nemeschansky \cite{MN1,MN2}, we assume the SW
differential in $\rep$ to take the form
\beq
\lambda_\rep =\left( c_1z+c_3 B(w)\right) {dx \over y}
+c_2\sum_a {m_a(w)y_a(z) \over x-x_a(z)} {dx \over y},
\label{swdiff}
\eeq
where $B(w)=w_2$ for $D_4$, $w_2^2$ for $E_7$, $w_2^3$ for $E_8$ and $0$ 
otherwise, and constants $c_i$ will be determined up to the overall 
normalization in such a way that $\lambda_\rep$ obeys (\ref{cond}). 
Note that given the degree 1 to $m_a(w)$, $\lambda_\rep$ has 
the degree 1 which equals mass 
dimension of $\lambda_\rep$. Since $m_a=(\lambda_a ,\phi)$ as in (\ref{zero}),
the $\phi_i$ are $r$ $(={\rm rank}\; G)$ independent mass parameters in the
theory.

In the following we construct $\lambda_\rep$ 
explicitly for the $A_1$, $A_2$, $D_4$ and $E_n$ $(n=6,7,8)$ theories. 
The $A_1$ case is too simple to exhibit the essence of our
calculations. So we start with the case of $A_2$ which is not only instructive
but tractable by hand. In the $D_4$ and $E_n$ theories 
we have used the Maple
software on computer to carry out our calculations. The $A_1$ result is
given at the end of this section. A full detail of how to evaluate 
$\pa \lambda_\rep /\pa z$ is presented in Appendix A. 
The data of characteristic 
polynomials for $D_4,\; E_6,\; E_7$ and $E_8$ is collected in Appendix B.

\subsection{The $A_2$ theory}

The $A_2$ curve is written in terms of the coefficient polynomials 
(\ref{a2curve}). As a section let us assume
\beq
x=v, \hskip10mm y=z.
\label{a2f}
\eeq
with $v \in {\bf C}$. Substituting this into the $A_2$ curve
it is obvious that $v$ has to satisfy 
\beq
v^3+w_2v+w_3=0.
\label{vcubic}
\eeq
The LHS is in the form of the characteristic polynomial $P^{\bf 3}_{A_2}(t)$ 
for ${\bf 3}$ of $SU(3)$ with two Casimirs $w_2$ and $w_3$ under the relation
$t \propto v$. Thus $v$ is determined by the three zeroes $m_a$ of 
$P^{\bf 3}_{A_2}(t)$. Let us set $t=v/2$,\footnote{There is no a priori
reason for fixing a constant $c$ in the relation $t=cv$. Our choice
$t=v/2$ will be justified in section 7 by considering the renormalization
group flows from ({\it or} to) the $D_4$ theory. 
This remark also applies to the
following cases studied in this section.} then we have the three roots $v_a$ of
(\ref{vcubic}) as $v_a=2m_a$ and
\beqa
&& w_2=v_1v_2+v_2v_3+v_3v_1=-4(\phi_1^2+\phi_2^2-\phi_1\phi_2), \CR
&& w_3=-v_1v_2v_3 = -8 \phi_1\phi_2(\phi_1-\phi_2)
\eeqa
with $v_1+v_2+v_3=0$. Putting $v=v_a$ we observe that the section (\ref{a2f})
belongs to ${\bf 3}$ of $SU(3)$. 

It is quite interesting that the characteristic polynomial naturally appears
when the global sections are determined. Accordingly the residues of the
differential $\lambda_\rep$ are fixed as was discussed before. We thus write
down $\lambda_{\bf 3}$ in the form
\beq
\lambda_{\bf 3}=c_1z{dx \over y}
+c_2\sum_{a=1}^3 {m_ay_a \over x-x_a}{dx \over y}, 
\hskip10mm m_a={v_a \over 2} .
\eeq
Note that the sum of the residues has to vanish. This is ensured since there
also exist poles with residues with opposite sign on the other sheet. These
poles belong to ${\bf \bar 3}$ of $SU(3)$. Here (\ref{dldz}) yields
\beq
{\pa \lambda_{\bf 3} \over \pa z}
={2c_1 \over 3} {dx \over y}+(A_1x+A_0){dx \over y^3}+d(*),
\eeq
where
\beq
A_1={1 \over 3}(2c_1-3c_2)w_2, \hskip10mm A_0=\half (2c_1-3c_2)w_3
\eeq
{}from which we get $c_1=3c_2/2$.

We can find another section by assuming
\beq
\left\{ \begin{array}{ll}
\displaystyle{\hskip1.5mm  x(z)={z^2 \over v^2}}+b_1z+b_0, & \\
\displaystyle{\hskip1.5mm  y(z)={z^3 \over v^3}}+r_2z^2+r_1z+r_0,  &
\end{array}
\right.
\label{adj}
\eeq
where $v, b_i, r_i \in {\bf C}$ . Plugging this in the $A_2$
curve one obtains the relations
\beqa
&& r_0^2-w_3-b_0^3-w_2b_0=0,   \CR
&& -3b_1b_0^2+2r_1r_0-w_2b_1=0, \CR
&& \left( 2r_2r_0+r_1^2-3b_1^2b_0-1 \right) v^2-3b_0^2-w_2 =0, \CR
&& 2r_0+\left( 2r_2r_1-b_1^3 \right) v^3-6b_1b_0v =0, \CR
&& -3b_0+2r_1v-3b_1^2v^2+r_2^2v^4=0, \CR
&& 2r_2v-3b_1=0.
\label{octrelation}
\eeqa
Eliminating $b_i$ and $r_i$ we are left with
\beq
64v^6+96w_2v^4+36w_2^2v^2+4w_2^3+27w_3^2=0,
\label{eq01}
\eeq
while the characteristic polynomial for ${\bf 8}$ of $SU(3)$ reads
\beq
P_{A_2}^{\bf 8}(t)=t^2 \left(t^6+{3 \over 2}w_2 t^4+{9 \over 16}w_2^2 t^2
+{w_2^3 \over 16}+{27 \over 64} w_3^2 \right) .
\label{eq02}
\eeq
Thus the six roots $v_a$ of (\ref{eq01}) are identified with the generically
non-vanishing zeroes of 
(\ref{eq02}), {\it i.e.} 
\beq
P_{A_2}^{\bf 8}(v_a)=0, 
\label{octzero}
\eeq
{}from which we see that
the section (\ref{adj}) belongs to ${\bf 8}$ of $SU(3)$.

For the adjoint section (\ref{adj}) the SW differential is constructed as
\beq
\lambda_{\bf 8}=c_1z{dx \over y}
+c_2\sum_{a=1}^3 {m_a y_a \over x-x_a}{dx \over y}, \hskip10mm   m_a=v_a.
\eeq
The non-zero weights of ${\bf 8}$ read $\lambda_{\pm 1}=\pm (1,1)$,
$\lambda_{\pm 2}=\pm (-1,2)$ and $\lambda_{\pm 3}=\pm (2,-1)$ in the Dynkin
basis. We have from (\ref{zero}) and (\ref{octzero}) that 
$v_{\pm a}=(\lambda_{\pm a}, \phi)$. Note that $v_{-a}=-v_a$ $(a=1,2,3)$
give the residues of the poles on the other sheet. In terms of this 
parametrization, one can find $v_ar_{1a}$ explicitly from (\ref{octrelation})
\beq
v_1r_{11}=-3 (\phi_1-\phi_2), \hskip10mm
v_2r_{12}=3 \phi_1 , \hskip10mm
v_3r_{13}=-3 \phi_2.
\eeq
$A_0$ and $A_1$ in (\ref{A1A0}) are then evaluated to be
\beq
A_1=(2 c_1-9c_2)w_2, \hskip10mm
A_0=-3 c_2z^2+ \half (2 c_1-9c_2)w_3.
\eeq
To manipulate the $z^2$ term in $A_0$ we note
\beq
z^2=W-{1 \over 3}x\pa_x W-{2\over 3}xf-(g-z^2)
\eeq
which yields
\beq
{z^2 \over W^{3/2}}={1\over 3\sqrt{W}}-
{1\over W^{3/2}}\left( {2w_2\over 3}x+w_3\right)
+{2 \over 3} \pa_x \left( {x \over \sqrt{W}}\right).
\eeq
Thus
\beq
{\pa \lambda_{\bf 8} \over \pa z}
=\left( {2c_1\over 3}-2c_2 \right) {dx \over y}
+ {2c_1 + 3c_2 \over 2} \left( {2w_2 \over 3}x+w_3 \right)
{dx \over y^3}+d(*),
\eeq
and hence we obtain $c_1=-3c_2/2$.

Finally it should be mentioned that the elliptic fibration (\ref{cubic}) with 
(\ref{e8curve})-(\ref{a1curve}) admits the section in
the form of (\ref{adj}) and, as we will see, (\ref{adj}) always
corresponds to the adjoint representation of $G=ADE$.

\subsection{The $D_4$ theory}

Taking the curve (\ref{d4curve}) for the $D_4$ theory we obtain the SW
differential in parallel with the $A_2$ case though the computations become
slightly more involved. Let us first examine the section in the form
\beq
\left\{ \begin{array}{ll}
\displaystyle{\hskip1.5mm  x(z)=\delta z+r,} & \\
\displaystyle{\hskip1.5mm  y(z)=v z+b.}  &
\end{array}
\right.
\label{d4vec}
\eeq
For $\delta=0$, plugging (\ref{d4vec}) in the $D_4$ curve gives
\beq
v^2-r-w_2=0, \hskip10mm 2vb-w_4=0, 
\hskip10mm b^2-r^3-r\widetilde w_4-w_6=0.
\eeq
The elimination procedure results in
\beq
v^8-3w_2v^6+\left( \widetilde w_4+3w_2^2 \right) v^4
+\left( w_6-w_2\widetilde w_4-w_2^3 \right) v^2 -w_4^2/4=0.
\label{d4elimi}
\eeq
This polynomial may be compared to the characteristic polynomial for
${\bf 8_v}$ (vector) of $SO(8)$
\beq
P^{\bf 8_v}_{D_4}(t)=t^8+u_2t^6+u_4t^4+u_6t^2-\widetilde u_4^2/4.
\eeq
Then (\ref{d4elimi}) is equivalent to
\beq
P^{\bf 8_v}_{D_4}(v_a)=0
\eeq
under the relation (\ref{shuffle}), showing that the section with $\delta =0$
is in the vector representation.

For $\delta=\pm i$, on the other hand, we observe
\beqa
&& P^{\bf 8_s}_{D_4}(v_a/2)=0,  \hskip10mm \hbox{for $\delta=+i$},  \CR
&& P^{\bf 8_c}_{D_4}(v_a/2)=0,  \hskip10mm \hbox{for $\delta=-i$}, 
\eeqa
where the characteristic polynomial for ${\bf 8_s}$ (spinor) of $SO(8)$
is given by
\beqa
&& P^{\bf 8_s}_{D_4}(t)
=t^8+u_2t^6+
\left( {3\over 8}u_2^2-{u_4\over 2}-{3i\over 2}\widetilde u_4 \right)t^4
+\left( -{u_2u_4 \over 4} +{u_2^3 \over 16}
-{i\over 4}u_2\widetilde u_4+u_6 \right)t^2  \CR
&& \hskip18mm 
-{u_2^2u_4 \over 32}-{i\over 8}\widetilde u_4u_4-{\widetilde u_4^2 \over 16}
+{i\over 32}u_2^2 \widetilde u_4+{u_4^2 \over 16}+{u_2^4 \over 256}
\eeqa
and that for ${\bf 8_c}$ (conjugate spinor) is obtained by replacing
$\widetilde u_4$ by $-\widetilde u_4$. Thus the sections with $\delta = \pm i$
are in the spinorial representations.

The SW differential for the ${\bf 8_v}$ section turns out to be
\beq
\lambda_{\bf 8_v}=c_1 z{dx \over y}
+{c_1 \over 2} \sum_{a=1}^4{m^v_a y_a \over x-x_a}{dx \over y} ,
\hskip10mm m^v_a =v_a, 
\label{diff8v}
\eeq
where $v_a=(\lambda_a, \phi)$ with $\lambda_1=(1,0,0,0)$, 
$\lambda_2=(-1,1,0,0)$,
$\lambda_3=(0,-1,1,1)$ and $\lambda_4=(0,0,-1,1)$ in the Dynkin basis, 
while for the ${\bf 8_s}$ and ${\bf 8_c}$ sections we obtain
\beq
\lambda_{\bf 8_s}=c_1  \left(z+{3i\over 2}w_2 \right) {dx \over y}
-{c_1 \over 2} \sum_{a=1}^4{m^s_a y_a \over x-x_a}{dx \over y} ,
\hskip10mm  m^s_a ={v_a \over 2} ,
\label{diff8s}
\eeq
where $v_a=2(\lambda_a, \phi)$ with $\lambda_1=(0,0,0,1)$,
$\lambda_2=(0,1,0,-1)$, $\lambda_3=(1,-1,1,0)$ and $\lambda_4=(-1,0,1,0)$, and
\beq
\lambda_{\bf 8_c}=c_1 \left(z-{3i\over 2}w_2 \right) {dx \over y}
-{c_1 \over 2} \sum_{a=1}^4{m^c_a y_a \over x-x_a}{dx \over y} ,
\hskip10mm  m^c_a ={v_a \over 2} ,
\eeq
where $v_a=2(\lambda_a, \phi)$ with $\lambda_1=(0,0,1,0)$, 
$\lambda_2=(0,1,-1,0)$, $\lambda_3=(1,-1,0,1)$ and $\lambda_4=(-1,0,0,1)$.
These SW differentials obey
\beq
{\pa \lambda_\rep \over \pa z}={c_1 \over 2}{dx \over y}+d(*)
\eeq
for $\rep ={\bf 8_v},{\bf 8_s}$ and ${\bf 8_c}$.

As in the $A_2$ theory, (\ref{adj}) gives the section in ${\bf 28}$ (adjoint)
of $SO(8)$. After $b_i$ and $r_i$ are eliminated from the relations like
(\ref{octrelation}), $v$ is determined as
the 24 non-zero roots $\pm v_a$ $(a=1,\cdots ,12)$ of
\beq
P^{\bf 28}_{D_4}(\pm v_a)=0.
\eeq
Assuming the SW differential in the form
\beq
\lambda_{\bf 28}=(c_1z+c_3 w_2 ){dx \over y}+
c_2  \sum_{a=1}^{12}{m_a y_a \over x-x_a}{dx \over y}, \hskip10mm m_a=v_a,
\label{diff28}
\eeq
we find $c_1=c_3=0$ and
\beq
{\pa \lambda_{\bf 28} \over \pa z}=-6c_2 {dx \over y}+d(*).
\eeq
Thus there is no holomorphic piece in $\lambda_{\bf 28}$ \cite{MN2}.

Finally we derive the differential $\lambda_{SW}$
for the original SW curve (\ref{swd4}) in the
$D_4$ theory. For this let us first take $\lambda_{\bf 8_v}$ and
make a change of variables (\ref{xtoX})
\beq
 x=-{X \over \alpha \beta}+{u_2 \over 3}, \hskip10mm 
y={2 Y \over \alpha \beta (\alpha -\beta)}, \hskip10mm
 z={2 Z \over i(\alpha -\beta)}
+{i(\alpha +\beta)\over \alpha \beta (\alpha -\beta)}X.
\eeq
Since
\beq
\pa_x=-{\alpha +\beta \over 2}\pa_Z- \alpha\beta \pa_X,
\eeq
one has to take care of the total derivative term in 
$\pa \lambda_{\bf 8_v}/\pa z$ (see (\ref{dldz})) when converting 
$\lambda_{\bf 8_v}$ into $\lambda_{SW}^{\bf 8_v}$. The result reads
\beq
\lambda_{SW}^{\bf 8_v}
=c_1 \left( 2Z-{\alpha +\beta \over 2}u_2 \right){dX \over Y}
-i c_1 \sum_{a=1}^4 {m^v_aY_a^v \over X-X_a^v}{dX \over Y},
\label{SWdiff8v}
\eeq
where $X_a^v=-\alpha\beta (m_a^v)^2$ and $Y_a^v=[Y]_{X=X_a^v}$.
In a similar vein we obtain from $\lambda_{\bf 8_s}$ and 
$\lambda_{\bf 8_c}$ that
\beqa
&& \lambda_{SW}^{\bf 8_s}
=c_1 \left( 2Z+{\alpha -\beta \over 2} u_2 \right){dX \over Y}
+i c_1 \sum_{a=1}^4 {m_a^s Y_a^s \over X-X_a^s}{dX \over Y},  \CR
&& \lambda_{SW}^{\bf 8_c}
=c_1 \left( 2Z-{\alpha - \beta \over 2} u_2 \right){dX \over Y}
+i c_1 \sum_{a=1}^4 {m_a^c Y_a^c \over X-X_a^c}{dX \over Y} .
\eeqa
These differentials obey
\beq
{\pa \lambda_{SW}^{\rep} \over \pa Z}=c_1 {dX \over Y} +d(*)
\eeq
for $\rep ={\bf 8_v},{\bf 8_s}$ and ${\bf 8_c}$. 
Thus we set 
\beq
c_1={\sqrt{2} \over 8\pi}
\label{SWnorm}
\eeq
according to the normalization adopted in \cite{SW2}.

\subsection{The $E_6$ theory}

The global section which transforms in ${\bf 27}$ of $E_6$ is given by 
\beq
\left\{ \begin{array}{ll}
\displaystyle{\hskip1.5mm  x_a(z)=v_a z+b_a,} & \\
\displaystyle{\hskip1.5mm  y_a(z)=z^2+r_a z+s_a}  &
\end{array}
\right.
\label{e6fund}
\eeq
with $a=1,\cdots ,27$ \cite{Shio}. In fact, the elimination procedure yields
\beq
P^{\bf 27}_{E_6}(v_a)=0.
\eeq
This reflects the well-known fact in classical algebraic geometry
that the cubic surface in ${\bf P}^3$ contains exactly 27 lines \cite{Ha}.

The SW differential associated with the ${\bf 27}$ section is obtained as
\beq
\lambda_{\bf 27}=36c_2  z{dx \over y}+
c_2 \sum_{a=1}^{27}{m_a y_a \over x-x_a}{dx \over y} , \hskip10mm m_a=v_a,
\label{e6diff27}
\eeq
where the poles with opposite residues on the other sheet transform in the
${\bf \overline{27}}$ of $E_6$. Upon taking the derivative one gets
\beq
{\pa \lambda_{\bf 27} \over \pa z}= 12c_2 {dx \over y}+d(*).
\label{e6deriv}
\eeq

In the $E_6$ theory too, (\ref{adj}) yields the section in ${\bf 78}$ 
(adjoint) of $E_6$. We see that $v$ takes the values 
$\pm v_a$ $(a=1,\cdots ,36)$ which correspond to the 72 non-zero roots of
\beq
P^{\bf 78}_{E_6}(\pm 2 v_a)=0.
\eeq
Assuming the SW differential in the form
\beq
\lambda_{\bf 78}=c_1z{dx \over y}+
c_2  \sum_{a=1}^{36}{m_a y_a \over x-x_a}{dx \over y}, \hskip10mm m_a=2v_a,
\eeq
we find $c_1=0$ and
\beq
{\pa \lambda_{\bf 78} \over \pa z}=-24 c_2 {dx \over y}+d(*).
\eeq
As in the case of $D_4$ the holomorphic piece is absent in 
$\lambda_{\bf 78}$ \cite{MN2}.

\subsection{The $E_7$ theory}

The global section in ${\bf 56}$ of $E_7$ is obtained by taking \cite{Shio}
\beq
\left\{ \begin{array}{ll}
\displaystyle{\hskip1.5mm  x(z)=c z+b,} & \\
\displaystyle{\hskip1.5mm  y(z)=v z^2+r z+s.}  &
\end{array}
\right.
\label{e7fund}
\eeq
We find after the elimination process that $v$ is determined from the 56
non-zero roots $\pm 2v_a$ ($a=1,\cdots ,56$) of
\beq
P^{\bf 56}_{E_7}(\pm 2v_a)=0.
\eeq
The SW differential associated with the ${\bf 56}$ section turns out to be
\beq
\lambda_{\bf 56}=48 c_2 (z+w_2^2){dx \over y}+
c_2 \sum_{a=1}^{28}{m_a y_a \over x-x_a}{dx \over y} , \hskip10mm m_a=2 v_a,
\eeq
{}from which we get
\beq
{\pa \lambda_{\bf 56} \over \pa z}= 12 c_2 {dx \over y}+d(*).
\label{e7-56}
\eeq

The section given by (\ref{adj}) again corresponds to ${\bf 133}$ 
(adjoint) of $E_7$. We see that $v$ takes the values 
$\pm v_a$ $(a=1,\cdots ,63)$ which yield the 126 non-zero roots of
\beq
P^{\bf 133}_{E_7}(\pm 2v_a)=0.
\eeq
We obtain the SW differential as
\beq
\lambda_{\bf 133}=-18 c_2 z{dx \over y}+
c_2  \sum_{a=1}^{63}{m_a y_a \over x-x_a}{dx \over y}, \hskip10mm m_a=2v_a,
\eeq
and
\beq
{\pa \lambda_{\bf 133} \over \pa z}=-36 c_2 {dx \over y}+d(*).
\label{e7-133}
\eeq

\subsection{The $E_8$ theory}

By counting degrees it is seen that there are no sections in the form of
(\ref{e6fund}), (\ref{e7fund}). This distinguishes the $E_8$ case from
$E_6$ and $E_7$, and corresponds to the fact that the fundamental of $E_8$ is 
identical with the adjoint. It is indeed proved by the elimination
procedure that the $E_8$ curve possesses the section as in 
(\ref{adj}) which transforms in ${\bf 248}$ of $E_8$ \cite{Shio}.
As explained in \cite{Shio}, one can explicitly evaluate the resultant which
appears in the final step of the elimination process. 
The result is that $v$ takes the values 
$\pm v_a$ $(a=1,\cdots ,120)$ which give the 240 non-zero roots of
\beq
P^{\bf 248}_{E_8}(\pm 2v_a)=0.
\eeq
The SW differential in ${\bf 248}$ is then found to be
\beq
\lambda_{\bf 248}=-2 c_2 (60z +w_2^3) {dx \over y}+
c_2  \sum_{a=1}^{120}{m_a y_a \over x-x_a}{dx \over y}, \hskip10mm m_a=2v_a
\eeq
and
\beq
{\pa \lambda_{\bf 248} \over \pa z}=-60 c_2 {dx \over y}+d(*).
\eeq

\subsection{The $A_1$ theory}

It is clear that the $A_1$ curve admits the section
\beq
x=0, \hskip10mm y=v
\eeq
which transforms in ${\bf 2}$ of $SU(2)$ since $v^2-w_2=0$. 
The SW differential in ${\bf 2}$ is easily obtained as
\beq
\lambda_{\bf 2}=c_2 \left( {z \over 3}
+{m_1y_1 \over x}\right) {dx \over y},
\eeq
where $m_1=v_1/2=\sqrt{w_2}/2$ and $y_1=v_1$. Thus we have
\beq
\lambda_{\bf 2}={c_2 \over 2} 
\left( {2 z \over 3}+{w_2 \over x}\right) {dx \over y}
\eeq
which obeys
\beq
{\pa \lambda_{\bf 2} \over \pa z}={c_2 \over 4} {dx \over y}+d(*).
\eeq

The section in ${\bf 3}$ of $SU(2)$ is given by
\beq
x={z^2 \over v^2}, \hskip10mm y={z^3 \over v^3}+{v \over 2}
\eeq
as in (\ref{adj}). Here $v$ satisfies $v^2-4w_2=0$ while 
$P_{A_1}^{\bf 3}(t)=t(t^2-w_2)$, and hence ${\bf 3}$ is realized.
Correspondingly we find 
\beqa
\lambda_{\bf 3} &=& c_2 \left( -{5z \over 6}+{m_1y_1 \over x-x_1}\right)
{dx \over y}  \CR
&=& {c_2 \over 2} \left( -{5z \over 3}
+ {z^3+8w_2^2 \over 4w_2 x-z^2} \right) {dx \over y},
\eeqa
where $m_1=v_1/2 =\sqrt{w_2}$. This differential obeys
\beq
{\pa \lambda_{\bf 3} \over \pa z}=- c_2 {dx \over y}+d(*).
\eeq

\section{The scaling limit}

\renewcommand{\theequation}{6.\arabic{equation}}\setcounter{equation}{0}

According to the results in the previous section, it is inferred 
that one can always
construct the SW differential $\lambda$ in the fundamental as well as 
adjoint representations in general $ADE$ case. For $D_4$, moreover,
we have obtained $\lambda_{\bf 8_\bullet}^{D_4}$ 
for the vector, spinor and conjugate spinor
of $SO(8)$ which are permuted under the triality automorphism of $D_4$.
Thus there arises a natural question whether the physics depends on
representations chosen in constructing the SW differential. 
In order to study this problem it is important to analyze how the SW 
differential behaves under the renormalization group flow. 

Let us analyze in great detail how the $E_6$ SW differential
reduces to the $D_4$ SW differential when we move simultaneously $\bA$- and
$\bC$-branes out to infinity from the $E_6$ seven brane background.
When a $\bA$-brane is removed the $E_6$ symmetry
breaks down to $SO(10) \times U(1)$.
The $E_6$ mass parameters $\phi_i$ are decomposed under the 
$SO(10) \times U(1)$ subgroup as
\beqa
&& \phi_1=2 M_1+b_1, \hskip10mm   \phi_2=4 M_1+b_2, 
\hskip10mm   \phi_3=6 M_1+b_3, \CR
&& \phi_4=5 M_1+b_4, \hskip10mm   \phi_5=4 M_1, 
\hskip19mm   \phi_6=3 M_1+b_5,
\eeqa
where the $b_i$ are the $SO(10)$ mass parameters and $M_1$ is the $U(1)$ mass
\cite{TeYa4}. Here the mass parameters are labeled as shown in the Dynkin
diagrams (see Fig.1). 
\begin{figure}
\hspace{1.5cm}
\epsffile{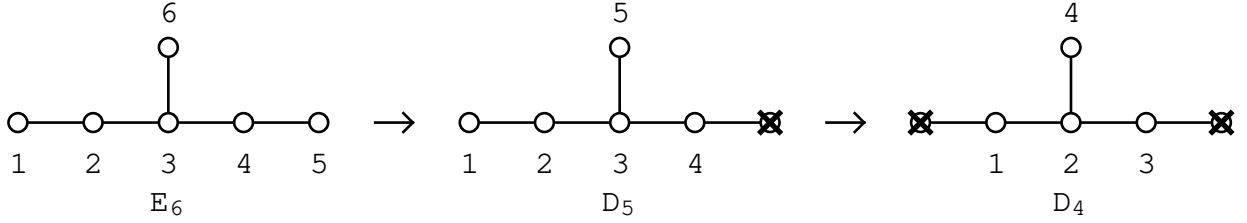}
\caption{Dynkin diagrams}
\label{fig:E6D5D4}
\end{figure}
Removing a $\bC$-brane induces the breaking of 
$SO(10)$ to $SO(8) \times U(1)$. Under $SO(8) \times U(1)$ the $SO(10)$ mass
parameters are decomposed into the $SO(8)$ masses $c_i$ and $U(1)$ mass $M_2$
as follows:
\beqa
&& b_1=M_2, \hskip10mm b_2=M_2+c_1, \hskip10mm b_3=M_2+c_2, \CR
&& b_4=M_2/2+c_3, \hskip10mm   b_5=M_2/2+c_4.
\eeqa
Upon sending $\bA$- and $\bC$-branes together to infinity we take the
scaling limit \cite{MN1}
\beq
M_1, \; M_2 \rightarrow \infty, 
\hskip10mm {M_1 \over M_2}= -{\A+\B \over 6(\A-\B)}\quad {\rm fixed},
\label{slimit}
\eeq
where the limit $M_i \rightarrow \infty$ decouples two $U(1)$ factors
and the ratio with $\alpha ,\, \beta$ defined in (\ref{d4param}) gives 
the value of the marginal gauge coupling constant in the $D_4$ theory.

In order to see that the $E_6$ curve reduces to 
the $SO(8)$ SW curve (\ref{swd4}) we first write the $E_6$ invariants 
$w_{q_i}(\phi)$ in terms of $SO(8)$ masses $c_i$ 
\beqa
&& \phi_1 = \frac{4}{3} (\A -2 \B) M,  \hskip7mm 
\phi_2= \frac{2}{3} ( \A -5 \B) M+c_1,   \hskip7mm 
\phi_3= -4 \B M+c_2, \CR
&& \phi_4= -\frac{2}{3} (\A +4 \B) +c_3,   \hskip7mm
\phi_5= -\frac{4}{3} (\B+\A) M,  \hskip7mm
\phi_6= - 2 \B M+c_4, 
\label{atoc}
\eeqa
where $M=-3 M_1 /(\A+\B)=M_2/( 2 (\A-\B) )$ and the explicit expressions
of $w_{q_i}(\phi)$ in terms of $\phi_i$ are given in \cite{TeYa3} . 
Then making a change of variables 
\beqa
y & =& -i M^3 Y,     \CR
x & =& M^2 \left( -X-\frac{1}{12} (\A-\B)^2 u_2
+\frac{1}{3} (\A+\B) Z \right) ,    \CR
z &=& \frac{2}{27}( \B-2 \A)(\A-2 \B)(\A+\B) M^3
+M \left(  -\frac{1}{2} Z+\frac{1}{24} (\A+\B) u_2 \right)
\label{coodetod}
\eeqa
in the $E_6$ curve and letting $M \rightarrow \infty$, we obtain the
$SO(8)$ curve (\ref{swd4}) where $m_a =m_a^v=(\lambda_a, c)$ with $\lambda_a$
being a weight vector of ${\bf 8_v}$ in section 5.2. 

We next show explicitly that, in the limit (\ref{slimit}), the $E_6$ SW 
differential $\lambda_{\bf 27}^{E_6}$ in ${\bf 27}$ is reduced to 
the sum of the $D_4$ SW differentials
in ${\bf 8_v}, {\bf 8_s}, {\bf 8_c}$ we have
constructed previously. This corresponds to the fact that the fundamental 
representation ${\bf 27}$ of $E_6$ is decomposed under the $SO(8)$
subgroup into 
\beq
{\bf 27}={\bf 8_v}\oplus {\bf 8_s}\oplus {\bf 8_c}\oplus {\bf 1}
\oplus {\bf 1} \oplus {\bf 1}.
\label{27to8s}
\eeq
Let us put (\ref{atoc}), (\ref{coodetod}) in the $E_6$ differential
(\ref{e6diff27})
\beq
\lm^{E_6}_{\bf 27}=c_2
\left( 36 z + \sum_{a=1}^{27}
\frac{m_a(\phi_i) y_a(z,\phi_i)}{x-x_a(z,\phi_i)} \right)
\frac{dx}{y}.
\eeq
and let $M \rightarrow \infty$, then we obtain
\beq
36 z \frac{1}{i} \frac{dx}{y} =\left( 
-\frac{3}{8} (\A+\B)(2 \A-\B)(2\B-\A) M^2 
+18 u -\frac{3}{2} (\A+\B) u_2(c_i) 
\right) \frac{dX}{Y} +O \left( \frac{1}{M} \right).
\label{dxy}
\eeq
The poles in the singlets of $SO(8)$ go to infinity in this limit.
Remember that the poles appear pairwise on two sheets of the Riemann surface
in such a way that the sum of residues vanish. Indeed we have 
\beqa
&&  \hskip5mm
\sum_{a \in {\cal S}} \frac{1}{i} \frac{m_a y_a}{x-x_a} \frac{dx}{y} \CR
&& = \left(  \frac{3}{8} (\A+\B)(2 \A-\B)(2\B-\A) M^2 
-2 Z+\frac{1}{6} (\A+\B) u_2(c_i) \right) 
\frac{dX}{Y} +O \left( \frac{1}{M} \right),
\label{sing}
\eeqa
where ${\cal S}$ denotes a set of $SO(8)$ singlets, and hence
the divergent pieces of (\ref{dxy}) and (\ref{sing}) cancel out.

The pole terms in ${\bf 8_v}$ turn out to be
\beq
\frac{1}{i} \frac{m_a y_a}{x-x_a} \frac{dx}{y}=
M \frac{A^{a}_1(Z,c_i)+A^{a}_2(Z,c_i) \frac{1}{M} 
+O \left( \frac{1}{M} \right)}{X-A^{a}_3(Z,c_i)
+A^{a}_4(Z,c_i)\frac{1}{M} +O \left( \frac{1}{M} \right)} \frac{dX}{Y},
\label{aaa1}
\eeq
where $A^{a}_i$ is a polynomial of $Z$ and $c_j$. Although this
seems to be divergent at first sight,
the poles associated with weights $\lambda$ and $-\lambda$ in ${\bf 8_v}$
coalesce at the same point, making these contributions finite
in the limit $M \rightarrow \infty$. It is verified that the
sum over terms with these weights $\pm \lambda$ of ${\bf 8_v}$ becomes 
finite,
\beqa
&& M \frac{A^{a}_1 +A^{a}_2  \frac{1}{M} +O \left( \frac{1}{M} \right)}{
X-A^{a}_3 +A^{a}_4 \frac{1}{M} +O \left( \frac{1}{M} \right)} \frac{dX}{Y}+
M \frac{-A^{a}_1 +A^{a}_2  \frac{1}{M} +O \left( \frac{1}{M} \right)}{
X-A^{a}_3 -A^{a}_4 \frac{1}{M} +O \left( \frac{1}{M} \right)} 
\frac{dX}{Y} \CR
&=& \frac{-2 A^{a}_1 A^{a}_4}{(X-A^{a}_3)^2} \frac{dX}{Y} +
\frac{2 A^{a}_2}{X-A^{a}_3} \frac{dX}{Y}+O \left( \frac{1}{M} \right),
\label{aaa2}
\eeqa
where we found that $A^a_3$ is equal to the pole position
$X_a^v$ of $\lambda_{\bf 8_v}^{D_4}$. Thus we get 
\beqa
\sum_{a \in {\bf 8_v}} \frac{1}{i} \frac{m_a y_a}{x-x_a} \frac{dx}{y}
& \rightarrow& 
\sum_{a =1}^{4} \left( 
\frac{-2 A^{a}_1 A^{a}_4}{(X-X^v_a)^2} \frac{dX}{Y} +
\frac{2 A^{a}_2}{X-X^v_a} \frac{dX}{Y} \right) \CR
&=&\sum_{a =1}^{4} \left( 
d \left(\frac{2 A^{a}_1 A^{a}_4}{X-X^v_a} \frac{1}{Y} \right)
+\frac{ A^{a}_1 A^{a}_4}{X-X^v_a} \frac{1}{Y^2} \frac{\pa Y^2}{\pa X} 
\frac{dX}{Y} +
\frac{2 A^{a}_2}{X-X^v_a} \frac{dX}{Y} \right) 
\CR
&=& \sum_{a =1}^{4} \left( 
\frac{1}{X-X^v_a} \frac{A^{a}_1 A^{a}_4 \frac{\pa Y^2}{\pa X} 
+2 A_2^a Y^2 }{Y^2} \frac{dX}{Y} + 
d \left(\frac{2 A^{a}_1 A^{a}_4}{X-X^v_a} \frac{1}{Y} \right)
\right) ,   \CR 
\label{8vres}
\eeqa
where the sum on the RHS is taken over half of the weights of ${\bf 8_v}$.
We can proceed further by showing that
\beq
A^{a}_1 (Z,c_i) A^{a}_4(Z,c_i) = 
- \gamma_v \, [ Y^2]_{X=X^v_a}(Z,c_i),
\eeq
and
\beq
\gamma_v \left[ \frac{\pa Y^2}{\pa X} \right]_{X=X^v_a}
-2 A_2^a=2 i m^v_a \,\, [ Y]_{X=X^v_a}(Z,c_i),
\eeq
where $\gamma_v=\frac{\A+\B}{3 \A \B}$.

Thus 
\beq
\sum_{a \in {\bf 8_v}} \frac{1}{i} \frac{m_a y_a}{x-x_a} \frac{dx}{y}
\rightarrow 
\sum_{a =1}^{4} \left( 
\frac{1}{i} \frac{2 m^v_a [ Y]_{X=X^v_a}}{X-X^v_a} \frac{dX}{Y} 
-d \left(\frac{2 \gamma_v [ Y^2]_{X=X^v_a}}{X-X^v_a} \frac{1}{Y} \right) -
R^v_a \right),
\label{8vres2}
\eeq
where
\beq
R^v_a=\frac{\gamma_v}{Y^2} \frac{[ Y^2]_{X=X^v_a} \frac{\pa Y^2}{\pa X}
-Y^2 \left[ \frac{\pa Y^2}{\pa X} \right]_{X=X^v_a}}{X-X^v_a} \frac{dX}{Y}.
\label{8vres3}
\eeq

For the pole terms in ${\bf 8_s}$ of $SO(8)$ we obtain the result as in
(\ref{8vres2}) except that we put
$\A \rightarrow -\A$ and $\B \rightarrow \B-\A$ in $\gamma_v$ 
in (\ref{8vres3}) 
in accordance with the triality transformation and replace
$X^v_a$ and $m_a^v$ by $X^s_a$ and $m^s_a$ for ${\bf 8_s}$ respectively.
Likewise, for the pole terms in ${\bf 8_c}$ we let
$X^v_a \rightarrow X^v_c$, $m_a^v \rightarrow m_a^c$ and
$\B \rightarrow -\B$ and $\A \rightarrow \A-\B$ in (\ref{8vres2}).

Finally we sum up the three pieces from ${\bf 8_v}, {\bf 8_s},{\bf 8_c}$.
In doing so, we observe that
\beq
\sum_{r=v,s,c} \sum_{a=1}^4 R^r_a
=P_1(Z,c_i) \frac{dX}{Y}  
+2 d \left( \frac{P_1(Z,c_i) X -P_2(Z,c_i)}{Y} \right),
\eeq
where 
\beqa
P_1 &=& 4 Z -\frac{1}{3}(\A+\B) u_2(c_i),  \CR
P_2 &=& -\frac{2}{3}(\A+\B) (\A^2-\A \B +\B^2) u_4(c_i) 
-2 i (\A-\B)(\A^2-\A \B +\B^2) \widetilde u_4(c_i)  \CR
&& \,\, +\frac{1}{12}(\A+\B)^3 {u_2(c_i)}^2
-\frac{4}{3} (\A^2-\A \B +\B^2) u_2(c_i) Z +\frac{4}{3} (\A+\B) Z^2.
\eeqa
As a result, we find in the scaling limit that the $E_6$ SW differential
in ${\bf 27}$ is reduced to the $SO(8)$ ones as
\beqa
\lm^{E_6}_{\bf 27} &\rightarrow& i c_2  \left( 
12 Z \frac{dX}{Y} -(\A+\B) u_2(c_i) \frac{dX}{Y} +2 
\sum_{r=v,s,c} \sum_{a=1}^4 \frac{1}{i} 
\frac{m^r_{a} [Y]_{X=X^r_a} }{X-X^r_a} 
\frac{dX}{Y} \right) +d(*) \CR
&=& 8\pi \sqrt{2}i c_2 \left( \lm^{\bf 8_v}_{SW} 
+ \lm^{\bf 8_s}_{SW} +\lm^{\bf 8_c}_{SW} \right)+d(*),
\label{e6tod4l}
\eeqa
where $\lambda^{\bf 8_\bullet}_{SW}$ has been normalized as in (\ref{SWnorm}).

We encounter here a somewhat curious situation; $\lambda_{\bf 27}^{E_6}$ does
not reduce to one of the $\lambda^{\bf 8_\bullet}_{SW}$, but the sum of
$\lambda^{\bf 8_\bullet}_{SW}$. In view of (\ref{27to8s}) and 
$SO(8)$ triality, on the one hand, (\ref{e6tod4l}) seems natural. 
Then one would say that picking up any one of 
$\lambda^{\bf 8_\bullet}_{SW}$ is sufficient to describe the physics. Note,
however, that the location of poles and their residues depend on 
${\bf 8_v}, {\bf 8_s},{\bf 8_c}$, and it is not so obvious if the
irrelevance of which ${\bf 8}$ of $SO(8)$ we choose to construct the SW
differential is really due to triality invariance which is inherent in $SO(8)$.
In addition to this, the SW differential $\lambda_{\bf 78}^{E_6}$ 
looks totally different from $\lambda_{\bf 27}^{E_6}$. 
This is also the case in the $D_4$ theory. In what follows we will study
if the representation chosen in constructing $\lambda$ is relevant to the
physics or not.

\section{Universality of Seiberg-Witten periods}

\renewcommand{\theequation}{7.\arabic{equation}}\setcounter{equation}{0}

Having derived (\ref{e6tod4l}), how do we fix the normalization constant $c_2$
for $\lambda_{\bf 27}^{E_6}$? Let us first point out that, under the 
renormalization group flows (\ref{flow}), the period integrals
\beq
\oint {\pa \lambda_\rep^G \over \pa z}
\label{smoothperi}
\eeq
exhibit the smooth limiting behavior at the generic points on the moduli
space. Then we obtain from (\ref{e6deriv}) and (\ref{coodetod}) that
$c_2 ={\sqrt{2} \over 48\pi i}$
for $\lambda_{\bf 27}^{E_6}$. Eq.(\ref{e6tod4l}) is written as
\beq
\lambda_{\bf 27}^{E_6} \rightarrow
{1 \over 3} \left( \lm^{\bf 8_v}_{SW}
+ \lm^{\bf 8_s}_{SW} +\lm^{\bf 8_c}_{SW} \right)+d(*).
\eeq
We also observe that the residues of the poles of 
$\lambda_{\bf 27}^{E_6}$ turn out to be
\beq
2 \pi i {\rm Res}_{x=x_a}(\lambda_{\bf 27}^{E_6})
=\frac{1}{k_{\bf 27}} \frac{m_a}{2 \sqrt{2} }
\eeq
with $k_{\bf 27}=6$. Notice that the index of ${\bf 27}$ 
(or ${\bf \overline {27}}$) is equal to 6. The appearance of the index of 
representations is not peculiar to this case. For example, in (\ref{e7-56})
and (\ref{e7-133}) we see $12=\ell ({\bf 56})$ and 
$36 =\ell ({\bf 133})$, respectively, where $\ell(\rep)$ is the index of the
representation $\rep$ (see Table 3). 

\begin{table}
\begin{center}
\begin{tabular}{cll} \hline
    & index     &    \\ \hline
$A_1$  & $ \ell({\bf 3})=4  $ & $ \ell({\bf 2})=1$    \\
$A_2$  & $  \ell({\bf 8})=6 $ & $ \ell({\bf 3})=\ell({\bf \bar 3})=1$  \\
$D_4$  &  $   \ell({\bf 28})=12 
       $ & $ \ell({\bf 8_v})=\ell({\bf 8_s})=\ell({\bf 8_c})=2   $  \\
$E_6$  &  $  \ell({\bf 78})=24 
       $ & $ \ell({\bf 27})=\ell({\bf \overline{27}})=6  $  \\
$E_7$  & $ \ell({\bf 133})=36  $  & $ \ell({\bf 56})=12 $   \\
$E_8$  & $ \ell({\bf 248}) =60   $ \\ \hline
\end{tabular}
\end{center}
\caption{Index of representations. $\ell$(adjoint)$=2h$ and $\ell({\bf 1})=0$.}
\label{tbl3}
\end{table}

Now examining the SW differentials obtained for various instances 
in section 5 and the renormalization group flows (see also 
\cite{MN2,SW2,APSW}), we find that the residue should be normalized as
\beq
2 \pi i {\rm Res}_{x=x_a}(\lambda_\rep^G) 
=\frac{1}{k_\rep } \frac{m_a}{2 \sqrt{2} }, \hskip10mm m_a=(\lambda_a, \phi),
\label{normres}
\eeq
where $k_\rep =\ell(\rep)$,
or $k_\rep =\ell(\rep)/2$ if the sum of the poles is taken over half of the
(non-zero) weights of $\rep$, and we use $\lambda_{SW}^\rep$ for the $D_4$
differential. Here the mass parameter $\phi$ is normalized
so that we have $m_a^v=(\lambda_a, \phi)$ in the $D_4$ theory along the
flows (\ref{flow}). This explains why we need to be a little careful to
fix a numerical constant upon relating $m_a$ and $v_a$ in section 5.
With this normalization of residues, it can be checked that the two-form
$\Omega$ in (\ref{2form}) is invariant under the successive
flows (\ref{flow}). We also see that
\beq
{\pa \lambda_\rep^G \over \pa z}= \kappa_G {dx \over y},
\label{indepnorm}
\eeq
where $\kappa_G$ is independent of $\rep$ as read off from section 5.

We claim that (\ref{normres}) is the correct normalization of the residue.
Taking this for granted, consider the renormalization group flow from 
the $G$ theory to the $G'$ theory.
In the $G$ theory, let the residues of the SW differential transform in the
representation $\rep$ of $G$. If the $\rep$ branches to $\oplus_i \rep'_i$
under $G \supset G'$, the pole terms of $\lambda^G_\rep$ reduce 
to the sum of pole terms
each of which transforms in $\rep'_i$ of $G'$. As observed in the flow
$E_6 \rightarrow D_4$ we expect that the pole terms belonging to non-singlets
of $G'$ remain finite in the scaling limit which implements the flow
$G \rightarrow G'$. If this is assumed to be the case, we obtain
\beq
\lambda^G_\rep \longrightarrow \sum_{\rep'_i \not= {\bf 1}}
{\ell(\rep'_i)\over \ell(\rep)} \lambda^{G'}_{\rep'_i}+d(*)
\label{lambdaflow}
\eeq
by matching the normalization of residues. This behavior is actually observed
in (\ref{e6tod4l}).

\subsection{Irrelevance of representations}

For a branching $\rep =\oplus_i \rep'_i$, we recall the identity
$\ell(\rep)=\sum_i \ell(\rep'_i)$.\footnote{This identity holds for the
regular embedding since the embedding index is unity. Every embedding in 
the flows (\ref{flow}), (\ref{flowD}) is regular.}  
Then (\ref{lambdaflow}) may imply
that the period integrals of $\lambda^{G'}_{\rep'_i}$ are independent 
of $\rep'_i$. This may sound surprising,
but we now prove that the SW differentials in any representation yield 
the identical physics result.

Since $\lambda^G_{\rep}$ has the poles with nonzero residues,
there is ambiguity in evaluating the periods
if we specify the cycles, along which $\lambda^G_{\rep}$ is integrated,
only in terms of the homology class of the SW curve.
Thus we consider the SW curve as the torus with punctures at 
the location of the poles of $\lambda^G_{\rep}$.
The homology class of this punctured torus
has a basis $\A$, $\B$ and $\gamma_a$.
Here $\gamma_a$ goes around a pole at $x=x_a$ counterclockwise, 
and the cycles $\A$ and $\B$ will be specified later.

Given $\lambda^G_{\rep_j}$ in the representation $\rep_j$, we define
\beq
a_{\rep_j}(z,\phi) = \oint_{\A} \lm^G_{\rep_j}, \hskip10mm
{a_D}_{\rep_j}(z,\phi)=\int_{\B} \lm^G_{\rep_j}
\eeq
and 
\beq
f(z,\phi) = a_{\rep_1} -a_{\rep_2} , \hskip10mm 
f_D(z,\phi) = {a_D}_{\rep_1} -{a_D}_{\rep_2}.
\eeq
It is an immediate consequence of (\ref{indepnorm}) 
that $f(z,m)=f(m)$ and $f_D(z,m)=f_D(m)$.
When we loop around a singularity at $z=z_k$ on the $z$-plane,
$\lambda_\rep^G$ remains invariant but
the cycles undergo the monodromy
\beqa
\A & \rightarrow& n \A+m \B+ \sum_a l_a \gamma_a, \CR
\B & \rightarrow& n' \A+m' \B+ \sum_a l'_a \gamma_a,
\eeqa
where the matrix $\pmatrix{n&m \cr n'&m'\cr}$ is conjugate to
$T=\pmatrix{1&1 \cr 0&1\cr}$ and $l,\, l'$ are some integers which are non-zero
when a cycle crosses a pole under a monodromy
transformation. At the singularity $z=z_k$, therefore, 
we have a linear relation
among $a_{\rep_j}$, ${a_D}_{\rep_j}$ and ${\rm Res}(\lambda^G_{\rep_j})$.
This in turn gives rise to a linear relation for $f(m)$, $f_D(m)$ and the
residues. A similar consideration at different singularity, say at $z=z_{k'}$,
 yields another linear relation. These two relations are linearly independent 
when two 7-branes at $z=z_k$ and at $z=z_{k'}$ are mutually non-local.
Then we find
\beq
f(m)= \sum_{j=1}^2 \sum_{a_j} c_{a_j}{\rm Res}_{x=x_{a_j}}(\lambda^G_{\rep_j}),
\hskip10mm
f_D(m)= \sum_{j=1}^2 \sum_{a_j} c'_{a_j}
{\rm Res}_{x=x_{a_j}}(\lambda^G_{\rep_j}),
\eeq
where $c_{a_j},\, c'_{a_j}$ are some constants. 
Hence we have shown that $f(m)$ and
$f_D(m)$ are linear in $m_a$. In fact, if $f(m)$ were not linear in $m_a$,
then for every $z$, we could have taken $1/ f(m) =0$ 
in the codimension one subspace of the space of bare mass parameters.
For a generic value of $z$, however, $f(m)$ may not be divergent,
and hence $f(m)$ should be linear in $m$.

Let us now apply a  Weyl transformations $m_a \rightarrow \widetilde m_a$
under which $\lm^G_{R_i}$ is left invariant. The SW periods $a(z,m)$ and
$a_D(z,m)$, however, may exhibit a non-trivial behavior under the Weyl
reflection. This occurs if the Weyl reflection moves a pole of 
$\lambda_\rep^G$ on the $x$-plane across the $\alpha$ and/or $\beta$ cycles.
The SW periods, on the other hand, should be Weyl invariant as gauge
invariant expectation values. We thus prescribe that the positions of the
cycles $\A$ and $\B$ are fixed relatively to the poles 
in such a way that the relative positions of the cycles
and the poles do not change under a Weyl transformation. Since it is always
possible to take such $\A$ and $\B$ in the asymptotic region
$z \gg m_a$ of the moduli space, we henceforth specify the cycles
according to this prescription.\footnote{See \cite{BF} for an explicit
example in the case of $N=2$ $SU(2)$ QCD with $N_f=2$ massive quarks.}
As a consequence of this, we see that $f(m)=f(\widetilde m)$ and
$f_D(m)=f_D(\widetilde m)$. Remember here the fact that 
there are no Weyl invariants which
are linear in $m_a$, and hence we obtain $f(m)=f_D(m)=0$. Therefore, we
conclude that the SW periods $a,\, a_D$ are independent of the choice of
a representation in constructing $\lambda_\rep^G$ as long as
the cycles $\A,\, \B$ are fixed properly as described above.

\subsection{Numerical check}

One can  numerically evaluate the period integrals and check that
$a$ and $a_D$ are independent of the 
representation $\rep$ of the residues of $\lm^G_\rep$.
In the case of $A_2$ and $D_4$, we express $a_\rep$ and ${a_D}_\rep$ for
$\rep =f$(undamental) and $adj$(oint) in terms of standard elliptic
integrals by taking two cycles $\A,\,\B$ as we prescribed above.
With the use of Maple, we then obtain, for example, 
$a_{f}= -28.99673387 + 16.74790178\, i$ 
and $a_{adj}= -28.99673386 + 16.74790178\, i$ in the $A_2$ theory at
$z=10$, $ m_1=1 -0.2\, i$ and $m_2=-0.4+0.75\, i$.
The error is indeed extremely small compared to the ratio of $z$ to $m_i$.
Varying the values of $m_i$ we plot in Fig.2 
the real and imaginary parts of $a_f$ in the $A_2$ theory 
for $z=10$ and $ m_1=2 x -0.2\, i,\; m_2=-0.4+1.5\, i\, x$.
Computing the periods at various values of $z$ and $m_i$,
we have observed in both $A_2$ and $D_4$ theories that
\beq
\frac{a_{f}-a_{adj}}{a_{f} }  <   10^{-8},   \hskip10mm
\frac{{a_D}_{f}-{a_D}_{adj}}{{a_D}_{f} }  <   10^{-8},
\label{dd}
\eeq
where the differentials (\ref{diff8v}), (\ref{diff28}) have been utilized
in the $D_4$ theory.
Since the values of $a$ and $a_D$ change substantially as shown in Fig.2
upon varying parameters of the moduli space,
we believe that the RHS of (\ref{dd}) are numerical errors and
really mean zero.

\begin{figure}
\begin{center}
\begin{tabular}{ll}
\ \epsfxsize=7cm \epsfbox{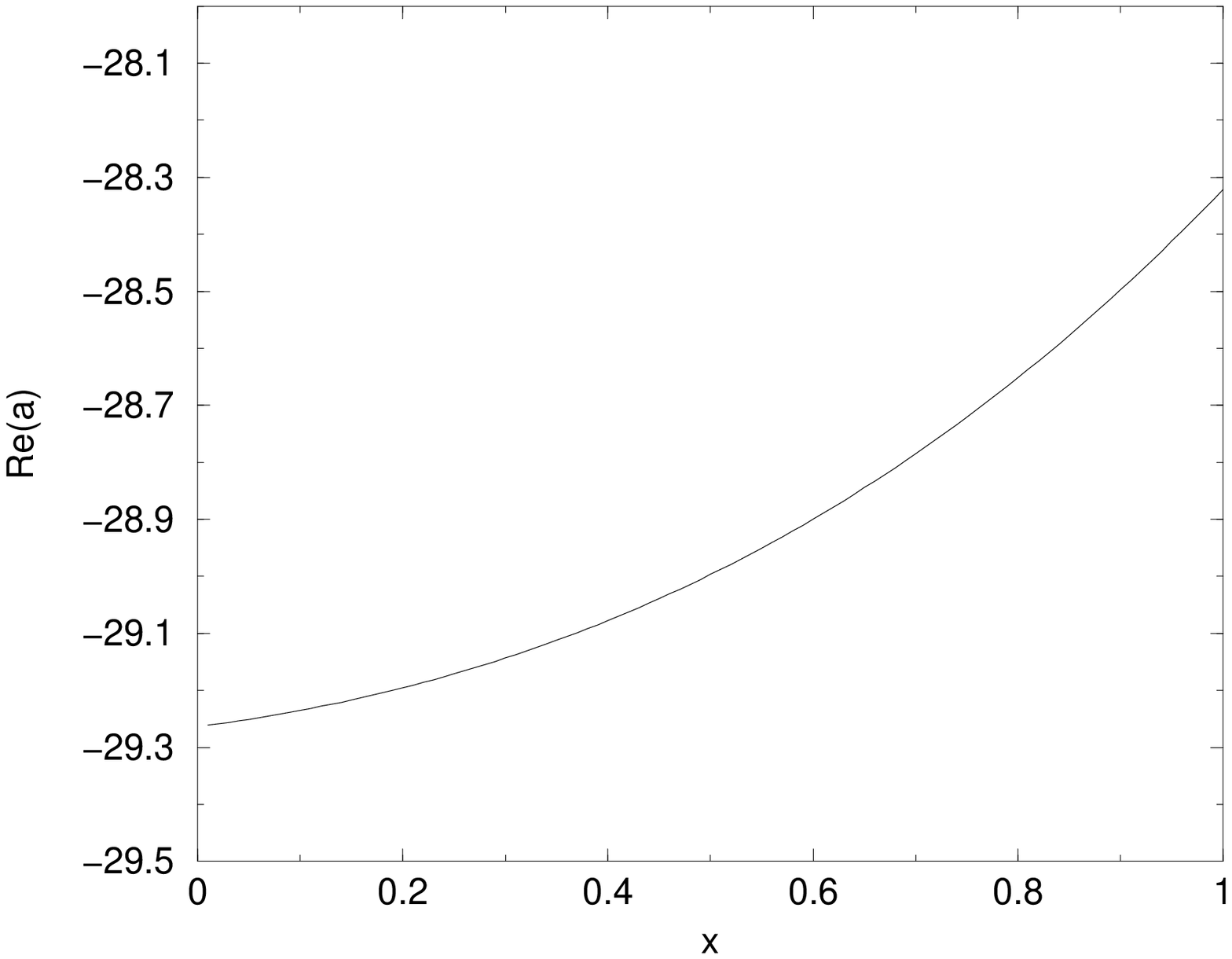} 
& \epsfxsize=7cm \epsfbox{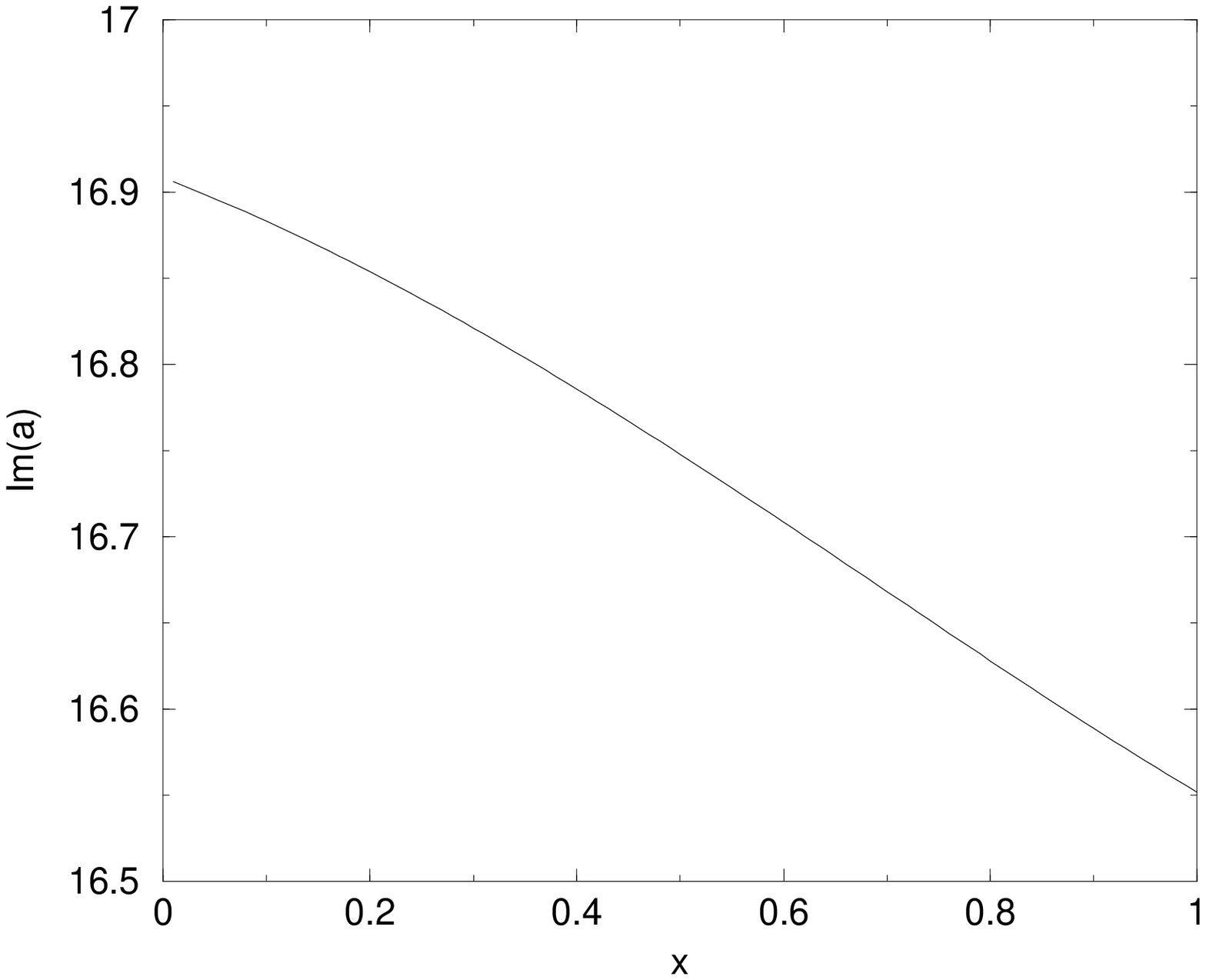}
\end{tabular}
\end{center}
\caption{The period $a=\int_\A \lm^{A_2}_{f}$ in the fundamental of $SU(3)$ 
is plotted for $z=10,\; m_1=2 x -0.2\, i, \; m_2=-0.4+1.5\, i\, x$ .}
\label{f1}
\end{figure}

To summarize, the SW periods will jump by a constant given by the residue
of $\lambda_\rep^G$ if we continuously deform the $\A,\, \B$ cycles across
a pole. Namely, the dependence of the periods on the cycles cannot be
absorbed by the redefinition of the periods among themselves. Furthermore,
since the positions of the poles and their residues are solely determined
by the representation $\rep$ chosen to construct $\lambda_\rep^G$, how to fix
the location of the cycles relatively to the poles is a subtle issue.
In spite of these, we have prescribed a way of specifying the cycles,
based on which the irrelevance of representations to the SW periods is proved.
To fix the BPS central charge, it remains to determine the constant piece
of the global abelian charges as mentioned in section 3. This will be
possible once we locate the cycles along which $\lambda_\rep^G$ is integrated.
It is also important to study the monodromy properties explicitly toward a
full account of the BPS spectrum.

\section{Flows to $N=2$ $SU(2)$ QCD with $N_f \leq 3$}

\renewcommand{\theequation}{8.\arabic{equation}}\setcounter{equation}{0}

It is well known that in $N=2$ $SU(2)$ QCD with $N_f$ fundamental quarks,
the global symmetry is enhanced to $SO(2N_f)$ when the quarks are 
massless \cite{SW2}. We now analyze how the SW differentials in the $N_f=4$
theory reduce to those in $N_f < 4$ theories. 

\subsection{Vector representation}

Let us first take the $N_f=4$ SW differential $\lambda_{SW}^{\bf 8_v}$ in 
the vector representation of $SO(8)$.
Upon taking the scaling limit  $\A \B \rightarrow 1,\; \A+\B \rightarrow -2$ 
and $m_4 \rightarrow \infty$ with $(\A-\B) m_4 = -\La_3/4$ fixed
\cite{SW2}, we obtain the $N_f=3$ theory.
In this limit the $D_4$ curve (\ref{swd4}), which can be rewritten as
\beq
Y^2=\A \B X \left( Z-\frac{(\A-\B) \A^2 \B^2 \prod_{b=1}^{4} m_b
+(\A+\B) X^2 }{2 \A \B  X} \right)^2
-\frac{(\A-\B)^2}{4 \A \B X} \prod_{a=1}^4 (X+\A \B m_a^2),
\eeq
becomes 
\beq
Y^2=X \left( X+Z+\frac{m_1 m_2 m_3 \La_3}{8 X} \right)^2
-\frac{\La^2}{64 X} \prod_{a=1}^3(X+m_a^2).
\eeq
This is shown to be equivalent to the usual $N_f=3$ curve
by setting $X=X'-Z$
\beqa
Y^2&=&X'^2 (X'-Z)- \frac{\La_3^2}{64} (X'-Z)^2
- \frac{\La_3^2}{64} ( m_1^2 + m_2^2 + m_3^2)  (X'-Z) \CR
&& +\frac{\La_3}{4} m_1 m_2 m_3 X' -\frac{\La_3^2}{64} 
(m_1^2 \, m_2^2+m_2^2 \, m_3^2+m_1^2 \, m_3^2).
\eeqa
Turning to the differential, one can verify that 
$\lm_{SW}^{ \bf 8_v}$ in (\ref{SWdiff8v}) with $X_a^v=- \A \B m_a^2$ and
\beq
Y_a=-i \A \B m_a \left( Z+ \frac{(\A-\B) \A^2\B^2 \prod_{a=1}^4 m_a 
+(\A+\B) \A^2 \B^2 m_a^4}{ 2 \A^2 \B^2 m_a^2}
\right),
\eeq
yields the $N_f=3$ SW differential 
\beq
\lm_{D_3}^{ \bf 6_v}=\frac{ \sqrt{2} }{8 \pi} 
\left( 2 Z-X'-(m_1^2+m_2^2+m_3^2) \right) \frac{dX'}{Y}
-\frac{ \sqrt{2} }{8 \pi} \sum_{a=1}^3 
\frac{m_a^2 Z-\frac{1}{8}m_1 m_2 m_3 \La_3 -m_a^4}{X'-Z+m_a^2} \frac{dX'}{Y}
\label{d3diff}
\eeq
which corresponds to the vector representation of $SO(6)$.

Taking here the limit $m_3 \rightarrow \infty$ with $\La_3 m_3=\La_2^2$ fixed,
we have the $N_f=2$ theory with the curve
\beq
Y^2=X'^2(X'-Z)- \frac{\La_2^2}{64} (X'-Z)+{\La_2^2 \over 4}m_1m_2X'
-{\La_2^4 \over 64}(m_1^2+m_2^2).
\label{nf2curve}
\eeq
The SW differential obtained from (\ref{d3diff}) turns out to be
\beq
\lm_{D_2}^{ \bf 4_v}=\frac{ \sqrt{2} }{8 \pi} 
\left( 2 Z-2 X'-(m_1^2+m_2^2) \right) \frac{dX'}{Y}
-\frac{ \sqrt{2} }{8 \pi} \sum_{a=1}^2 
\frac{m_a^2 Z-\frac{1}{8}m_1 m_2 \La_2^2 -m_a^4}{X'-Z+m_a^2} \frac{dX'}{Y}.
\eeq
Next, in the limit $m_2 \rightarrow \infty$ with $\La_2^2 m_2=\La_1^3$ fixed,
we obtain the $N_f=1$ curve from (\ref{nf2curve})
\beq
Y^2=X'^2(X'-Z)+{\La_1^3 \over 4}m_1X'-{\La_1^6 \over 64}
\eeq
and the differential
\beq
\lm_{D_1}^{ \bf 2_v}=\frac{ \sqrt{2} }{8 \pi} \left( 2 Z-3 X'-m_1^2 \right) 
\frac{dX'}{Y}
-\frac{ \sqrt{2} }{8 \pi} 
\frac{m_1^2 Z-\frac{1}{8}m_1 \La_1^3 -m_1^4}{X'-Z+m_1^2} \frac{dX'}{Y}.
\eeq
Finally, letting $m_1 \rightarrow \infty$ with $\La_1^3 m_1=\La_0^4$ fixed,
we arrive at the $N_f=0$ theory with the curve
\beq
Y^2=X'^2(X'-Z)+{\La_0^4 \over 4}X'
\eeq
and the standard form of the differential
\beq
\lm_{YM}=\frac{ \sqrt{2} }{8 \pi} \left( 2 Z-4 X' \right) \frac{dX'}{Y}.
\label{YM}
\eeq
Thus, under these renormalization group flows, we obtain the SW differentials
in the vector representation of $SO(2N_f)$.

We see from the above that the residues of $\lm_{D_n}^{ \bf {2n}_v}$ read
\beq
2\pi i {\rm Res} \left( \lm_{D_n}^{ \bf {2n}_v} \right)
={m_a \over 2\sqrt{2}}
\eeq
for $n\leq 4$, which agrees with (\ref{normres})
because $k_{ \bf {2n}_v}=1$, but differs from (17.1) 
of \cite{SW2}.\footnote{Our result resolves the puzzle 
in section 17 of \cite{SW2} 
why one has to replace $m_a$ by $m_a/2$ in the final form of the $N_f=4$ 
curve derived from the consideration of the residues. Thus it is also
required to make this replacement in (17.1) of \cite{SW2}, yielding the
correct result as we have obtained here.} 
The present result is the correct one since $\lm_{YM}$ derived through
the successive flows from $D_4$ coincides with that obtained in \cite{ItYa1}.
In order for this to hold, it is important that $\lm_{D_n}^{ \bf {2n}_v}$ 
obeys
\beq
\frac{\pa \lm_{D_n}^{ \bf {2n}_v}}{\pa Z}
=\frac{ \sqrt{2} }{8 \pi} \frac{dX'}{Y}.
\eeq
Furthermore it is clear that the massless limit of $\lm_{D_n}^{ \bf 2n_v}$ is 
in agreement with the ones obtained in \cite{ItYa1}.

\subsection{Spinor representation}

One may notice that the differentials $\lm_{D_n}^{ \bf 2n_v}$ do not look like
those obtained in \cite{SW2,AGMZ,BF}. Our next task is to show that they are
indeed derived from the $N_f=4$ SW differentials in spinors of $SO(8)$
and their residues transform in the spinor representation of $SO(2N_f)$
with $N_f \leq 3$.

First of all we note that the weights of ${\bf 8_s}$ of $SO(8)$ are given by
\beqa
&& m_1^s=\frac{1}{2} (m_1+m_2+m_3+m_4), \CR
&& m_2^s=\frac{1}{2} (m_1+m_2-m_3-m_4), \CR
&& m_3^s=\frac{1}{2} (m_1-m_2+m_3-m_4), \CR
&& m_4^s=\frac{1}{2} (m_1-m_2-m_3+m_4)
\eeqa
and $m_{4+a}^s=-m_a^s$. Note also 
$u_2=-\sum_{a=1}^4 m_a^2=-\sum_{a=1}^4 (m_a^s)^2$.
In $\lambda_{SW}^{\bf 8_s}$ (\ref{diff8s}), one has 
$X_a^s=\A Z -\A (\A-\B) ( \frac{1}{4} u_2 +( m_a^s)^2)$. Under the flow
$D_4 \rightarrow D_3$ generated by taking the scaling limit 
$m_4 \rightarrow \infty$, ${\bf 8_s}$ of $SO(8)$ is reduced to
${\bf 4_s}+{\bf 4_c}$ of $SO(6)$ where 
the weights of ${\bf 4_s}$ are 
\beqa
&& m_1^{\bf 4_s}=(m_1+m_2+m_3)/2, \hskip13mm  
m_2^{\bf 4_s}= -(m_1+m_2-m_3)/2 , \CR
&& m_3^{\bf 4_s}= -(m_1-m_2+m_3) /2, \hskip10mm   
m_4^{\bf 4_s}= (m_1-m_2-m_3)/2
\eeqa
and the weights of ${\bf 4_c}$ are $m_a^{\bf 4_c}=-m_a^{\bf 4_s}$. 
The positions $X_a^s$ of the poles become
\beqa
X_a^s &=&  -Z-\frac{1}{4} m_a^{\bf 4_s} \La_3  \CR
  &&  -\frac{1}{m_4} \Biggl( \frac{1}{32} m_a^{\bf 4_s} \La_3^2  
+ \frac{1}{8} Z \La_3 +\frac{1}{16} \La_3  \left(
-\sum_{b=1}^4 (m_b^{\bf 4_s})^2 +4 ( m_a^{\bf 4_s} )^2 \right) \Biggr)
+ O\left( \frac{1}{m_4^2} \right),  \CR
\eeqa
{}from which we see that the poles are not sent to infinity.
On the other hand, the residue is evidently divergent. This gives rise to a 
divergent piece in the SW periods in the scaling limit 
$m_4 \rightarrow \infty$. We note that this is a necessary divergence to make
certain BPS states decouple. To avoid this divergent behavior, though harmless,
let us alternatively take the differential
$\frac{1}{2} (\lm_{SW}^{ \bf 8_s}+\lm_{SW}^{ \bf 8_c})$. 
For this combination, we can evaluate the limit as performed in the flow from 
$E_6$ to $D_4$. The result is
\beq
\frac{1}{2} (\lm_{SW}^{ \bf 8_s}+\lm_{SW}^{ \bf 8_c}) \rightarrow
\lm_{D_3}^{ \bf 4_s} + dF(X',Z,m_a^{\bf 4_s}),
\eeq
where 
\beq
\lm_{D_3}^{ \bf 4_s}
=\frac{ \sqrt{2} }{8 \pi} \left( 2 Z-X' \right) \frac{dX'}{Y}
-\frac{ \sqrt{2} }{8 \pi} \sum_{a=1}^4 
\frac{m_a^{\bf 4_s} \frac{\La_3}{32} 
(4 Z-2 \sum_{b=1}^4 (m_b^{\bf 4_s})^2 +8 ( m_a^{\bf 4_s} )^2 
+\La_3 m_a^{\bf 4_s} )}{X'+\frac{1}{4} m_a^{\bf 4_s} \La_3} \frac{dX'}{Y}
\label{nf3diffs}
\eeq
and
\beqa
&& F= \frac{ \sqrt{2} }{256 \pi} \frac{1}{Y} \Biggl(
64 X'^2-64 (Z+\La_3^2) X'+16 m_1 m_2 m_3 \La_3    \CR
&&  \hskip30mm +2 Z \La_3^2 +64 \sum_{a=1}^4  
\frac{[Y]_{X'=-\frac{1}{4} m_a^{\bf 4_s}\La_3}}{X'
+\frac{1}{4}m_a^{\bf 4_s}\La_3} \Biggr) .
\eeqa
The differential (\ref{nf3diffs}) for the $N_f=3$ theory indeed agrees 
with \cite{AGMZ} and has the poles with residues in the form of (\ref{normres})
since the index of ${\bf 4_s}$ of $SO(6)$ is 1.

Next, taking the limit $ m_3 \rightarrow \infty$ with $\La_3 m_3=\La_2^2$ 
fixed to have the $N_f=2$ theory, it is shown that
\beq
\lm_{D_3}^{ \bf 4_s} \rightarrow
\lm_{D_2}^{{\bf 2_L}{\bf 2_R}} +
d \left(  \sqrt{2} \,
\frac{ 4 X'^2-4 Z X'+ m_1 m_2 \La_2^2 }{8 \pi Y}  
+ \frac{ \sqrt{2} }{2 \pi} \sum_{a=1}^2
\frac{[Y]_{X'=-\frac{1}{8} \La_2^2}}{X'+\frac{1}{8} \La_2^2} \frac{1}{Y}
\right),
\eeq
where
\beqa
\lm_{D_2}^{{\bf 2_L}{\bf 2_R}}
&=&\frac{ \sqrt{2} }{4 \pi} \left( Z- X' \right) \frac{dX'}{Y}
-\frac{ \sqrt{2} }{4 \pi} 
\left( \frac{m^{\bf 2_L} 
\frac{\La_2^2}{4}  m^{\bf 2_L} }{X'+\frac{1}{8} \La_2^2} 
-\frac{m^{\bf 2_R} \frac{\La_2^2}{4}  m^{\bf 2_R} }{X'-\frac{1}{8} \La_2^2} 
\right) \frac{dX'}{Y} \CR
&=&\frac{ -\sqrt{2} }{4 \pi} \frac{Y dX'}{X'^2-\frac{1}{64} \La_2^4},
\label{nf2diffsp}
\eeqa
which is in agreement with the one obtained in \cite{SW2}. 
Here ${\bf 4_s}$ of $SO(6)$
is decomposed into $({\bf 2},{\bf 1})+({\bf 1},{\bf 2})$ of 
$SU(2) \times SU(2)\, (=Spin(4))$ and the corresponding highest weights are
given by $m^{\bf 2_R}=(m_1+m_2)/2$ and $m^{\bf 2_L}=(m_1-m_2)/2$. Thus the
residues of $\lm_{D_2}^{{\bf 2_L}{\bf 2_R}}$ read off from (\ref{nf2diffsp})
become ${1 \over 2\pi i}{m_1\pm m_2 \over 2\sqrt{2}}$, which is the well-known
result \cite{SW2}. 

In the limit $ m_2 \rightarrow \infty$ with $\La_2^2 m_2=\La_1^3$ fixed,
we find the differential for the $N_f=1$ theory
\beq
\lm_{D_2}^{ \bf 2_s} \rightarrow
\lm_{D_1}^{ \bf 1_s} +
d \left(  \sqrt{2} \,
\frac{ - 2 X'^2+2 Z X'- m_1 \La_1^3-\La_1^6/x}{8 \pi Y}  \right),
\eeq
where
\beq
\lm_{D_1}^{ \bf 1_s}
=\frac{ \sqrt{2} }{8 \pi} \left( 2 Z- 3 X' \right) \frac{dX'}{Y}
-\frac{ \sqrt{2} }{8 \pi} 
\frac{m_1 \La_1^3}{4 X'} 
\frac{dX'}{Y}
\label{nf1diffsp}
\eeq
which again agrees with \cite{AGMZ,BF}.

Finally, we let $ m_1 \rightarrow \infty$ with $\La_1^3 m_1=\La_0^4$ fixed
to obtain the $N_f=0$ theory. In this limit we see that
the pole at $X'=0$ in (\ref{nf1diffsp}) turns out to be a double pole. Then,
using the $N_f=0$ curve $\frac{1}{X'}={1\over Y^2}
(X'^2-Z X'+\frac{1}{4} \La_0^4)$, we arrive at
\beq
\lm_{D_1}^{ \bf 1_s} \rightarrow
\lm_{YM}-\frac{ \sqrt{2} }{8 \pi} d \left( 
\frac{-4 X'^2+4 Z X'-\La_0^4}{2 Y} \right).
\eeq

In this section, we have shown   
that the SW differentials in the $N_f \leq 3$ theories can be
built from the vector as well as spinor representations of $SO(2N_f)$.
According to section 7 they describe the same physics in the Coulomb
branch of $N=2$ $SU(2)$ QCD with massive quarks. The SW differentials 
for $N_f \leq 3$ in general take the form
\beq
\lambda_{D_{N_f}}^\rep ={\sqrt{2} \over 8\pi}
\left( 2Z-(4-N_f)X' \right) {dX' \over Y}+(\hbox{pole terms}).
\eeq
Note here that $X'dX'/Y$ has double poles at infinity whose existence is
characteristic of the asymptotic freedom. It is interesting that
simple poles of $\lambda_{D_{N_f}}^\rep$ due to a massive quark become
congruent to the double poles at infinity in the scaling limit
$N_f \rightarrow N_f-1$.

\section{Conclusions}

In the framework of the F-theory compactification,
we have written down the elliptic curves for describing the $N=2$ theories with
$ADE$ global symmetries on a D3-brane in the Type IIB 7-brane background.
The SW differentials $\lambda$ have then been constructed for the fundamental
and adjoint representations of the $ADE$ groups. It is shown that the
physics results are independent of the representation of $\lambda$. It is
interesting to compare the present result with what has been known in
four-dimensional $N=2$ Yang-Mills theory with $ADE$ gauge symmetries. For
$N=2$ $ADE$ Yang-Mills theory the SW curves are given by the spectral curves 
whose form depends explicitly on the representations $\rep$ of $ADE$. However,
the physics of the Coulomb branch is equally described irrespective of $\rep$.
In \cite{MW} this is shown in terms of the universality of the special Prym 
variety known in the theory of spectral curves \cite{spect}. This is seen
more explicitly by analyzing the Picard-Fuchs equations for the SW
periods \cite{IY2}. Therefore, the universality we found here is considered
as the global symmetry version of the universality in $N=2$
Yang-Mills theory with local $ADE$ gauge symmetries.

It is clear in the framework of Type II string theory
that the $ADE$ global symmetries on a D3-brane and the $ADE$ gauge
symmetries of four-dimensional Yang-Mills theory have the common origin
in the $ADE$ singularities appearing in the degeneration of a $K3$ surface.
In fact, if we replace the top Casimir $w_h$ by 
$w_h+\rho +\Lambda^{2h}/\rho$ in (\ref{e8curve})-(\ref{a1curve}), 
our $ADE$ curves are recognized as the 
equations for the $ADE$ ALE space fibered over ${\bf P}^1$. Here $\rho$ is
a complex coordinate of the base ${\bf P}^1$. This reflects the 
compactification of Type II string theory on a $K3$ fibered Calabi-Yau
threefold. From this point of view, our calculation for the fundamental
of $E_6$ in section 5.3 is indeed equivalent to that in \cite{LW}
to obtain the SW curve for the $N=2$ $E_6$ Yang-Mills theory from the 
fibration of the $E_6$ ALE space. Hence our computations in section 5 
can be viewed
as the determination of the SW curves in the fundamental and adjoint 
representations for $N=2$ Yang-Mills theory with $ADE$ gauge symmetries.

The global sections of an elliptic fibration in higher representations
than the adjoint may be found by constructing the meromorphic 
sections.\footnote{We thank Y. Yamada for discussions on this point.}
The lattice structure hidden in our explicit computations will be related
to the lattice
which arises in the Mordell-Weil group. It will be interesting to
formulate our present results in more precise mathematical terms in view of 
the relation between the Mordell-Weil lattice and the $ADE$ singularity theory.

Finally, it is very interesting to analyze the BPS spectrum of the $E_n$
theories using our results. One application is to construct the 
junction lattice explicitly to describe the BPS states. This can be done at
least numerically as has been performed in $N=2$ $SU(2)$ 
theory \cite{BeFa,Oh}. In the $E_n$ theories the BPS spectrum possesses the
rich structure in comparison with the $D_{n \leq 4}$ theories \cite{dWHIZ}.
For instance, BPS states in arbitrary higher representations of the $E_n$ 
groups are shown to exist on the basis of (\ref{selection}). 
Combining the SW description properly formulated in the present paper and 
the junction approach will be efficient to gain a deeper understanding
of still mysterious four-dimensional $N=2$ superconformal field theories with
exceptional symmetry.

\vskip6mm\noindent
{\bf Acknowledgements}

\vskip2mm
We are grateful to K. Mohri, S. Sugimoto and Y. Yamada for useful discussions.
One of us (SKY) would like to thank T. Eguchi, 
J.A. Minahan and D. Nemeschansky for interesting conversations.
The research of ST is supported by JSPS Research Fellowship for Young 
Scientists. The research of MN and SKY was supported in part by Grant-in-Aid 
for Scientific Research on Priority Area 707 ``Supersymmetry and Unified
Theory of Elementary Particles'', 
Japan Ministry of Education, Science and Culture.

\newpage

\noindent
{\Large\bf Appendix A}

\renewcommand{\theequation}{A.\arabic{equation}}\setcounter{equation}{0}

\vskip3mm\noindent
We explain in detail how to evaluate $\pa \lambda_\rep /\pa z$. 
For an elliptic curve
\beq
y^2=W(x,z)
\label{appell}
\eeq
with
\beq
W(x,z)=x^3+f(z)x+g(z),
\label{Wform}
\eeq
the SW differential is assumed to be
\beq
\lambda_\rep =\left( c_1z+c_3B(w) \right) {dx \over y}
+c_2 \sum_a {v_a y_a(z) \over x-x_a(z)} {dx \over y}.
\eeq
Here $(x_a(z),y_a(z))$ are the global sections of an elliptic fibration
(\ref{appell}) and $v_a$ stand for the generically non-vanishing zeroes of the
characteristic polynomial for a representation $\rep$ of $G$
\beq
P_G^\rep (v_a)=0.
\eeq
Taking the derivative with respect to $z$, we obtain
\beqa
 {\pa \lambda_\rep \over \pa z}  
&=&  c_1 {2(q_z+q_x)-h \over 2q_z} {dx \over \sqrt {W}}
+{c_1 \over 2q_z W^{3/2}} {\cal L}W dx 
-{c_3B(w) \over 2 W^{3/2}}\left( x \pa_z f+\pa_z g \right) dx  \CR
&& +{c_2 \over 2 W^{3/2}}\sum_a 
\left( v_a(2\pa_zy_a W-y_a\pa_zW )
-v_a y_a \pa_z x_a\pa_x W \right){dx \over x-x_a} \CR
&& -\pa_x \left( {c_1q_x \over q_z}{x \over \sqrt {W}}          
 +c_2 \sum_a {v_a y_a \pa_z x_a\over (x-x_a) \sqrt {W}} \right) dx ,
\label{appdldz}
\eeqa
where we have defined the Euler operator
\beq
{\cal L}=\sum_iq_iw_{q_i}{\pa \over \pa w_{q_i}}
\eeq
in making use of the scaling equation for $W$
\beq
q_x x\pa_xW+q_z z \pa_zW+{\cal L}W=hW
\eeq
to rewrite the $z\pa_zW$ term. Notice that
\beqa
2\pa_z y_a &=& \pa_zW( x_a(z),z )   \CR
&=& \pa_zx_a [ \pa_xW ]_{x=x_a(z)}
+[ \pa_zW ]_{x=x_a(z)}.
\eeqa
Then the term with $dx/(x-x_a)$ in (\ref{appdldz}) is expressed as
\beq
{c_2 \over 2 W^{3/2}} \sum_a {v_a (B_1+B_2) \over x-x_a}{dx \over \sqrt{W_a}},
\eeq
where $W_a=W(x_a(z),z)$ and
\beqa
&& B_1=\pa_zx_a \left( W[ \pa_xW ]_{x=x_a(z)}-W_a\pa_xW \right), \CR
&& B_2=W [ \pa_zW ]_{x=x_a(z)}-W_a \pa_zW 
\eeqa
which vanish for $x=x_a(z)$. In fact, substituting (\ref{Wform}) one finds
\beqa
&& B_1=(x-x_a)\pa_zx_a 
\left( (x-x_a)^2f-3(x-x_a)g+3x_a^2x^2-6x_ag+f^2 \right), \CR
&& B_2=(x-x_a)\left( x_ax(x+x_a)\pa_zf+(x^2+x_ax+x_a^2)\pa_zg
+f\pa_zg-g\pa_zf \right) .
\eeqa
Now, after some algebra, we get
\beqa
{\pa \lambda_\rep \over \pa z}  
&=& c_1 {2(q_z+q_x)-h \over 2q_z} {dx \over \sqrt{W}}+\pa_x(\cdots ) dx \CR
&& +  \left( {c_1 \over 2q_z} (x {\cal L}f +{\cal L}g)
-{c_3 B(w)\over 2}(x\pa_z f+\pa_zg)+c_2(h_2x^2+h_1x+h_0) \right)
{dx \over W^{3/2}}, \CR
\eeqa
where
\beqa
&& h_2=\sum_av_a\pa_zy_a , \CR
&& h_1=\sum_av_a \left( x_a\pa_zy_a -{3\over 2}\pa_zx_ay_a \right) , \CR
&& h_0=\sum_av_a 
\left((x_a^2+f)\pa_zy_a-\half (\pa_zf+3x_a\pa_zx_a)y_a \right).
\eeqa
Using
\beq
{x^2 \over W^{3/2}}={1\over 3 W^{3/2}}(\pa_xW-f)= -{f \over 3 W^{3/2}}
-{2\over 3}\pa_x\left({1\over \sqrt{W}} \right) ,
\eeq
we arrive at
\beq
{\pa \lambda_\rep \over \pa z} =  {c_1 \over q_z} {dx \over y}
+\left( A_1(z) x+A_0(z) \right) {dx \over y^3}+\pa_x F(x,z)dx ,
\label{dldz}
\eeq
where we have used $q_x+q_z=q_y-1,\; 2q_y=h$, and
\beqa
&& A_1(z)={c_1 \over 2q_z} {\cal L}f-{c_3\over 2}B(w)\pa_zf+c_2h_1, \CR
&& A_0(z)={c_1 \over 2q_z} {\cal L}g-{c_3\over 2}B(w)\pa_zg
+c_2 \left( h_0-{1\over 3}h_2f \right),
\label{A1A0}
\eeqa
\beq
F(x,z)=-\left( {c_1q_x \over q_z} x+c_2 \sum_a{v_a y_a \pa_zx_a \over x-x_a}
+{2c_2h_2 \over 3}\right) {1 \over y}.
\eeq
At this stage one has to calculate $A_1,\, A_0$ which depend on the
explicit form of the section. After tedious calculations for higher rank
groups, the results are expressed in terms of
the deformation parameters $w_{q_i}$. Imposing $A_1=A_0=0$ now brings about the
overdetermined system with respect to $c_1$, $c_2$ and $c_3$. It is
quite impressive that we can nevertheless find the solution so as to
determine $c_i$ up to an overall normalization factor.

When we deal with the section in the adjoint representation (\ref{adj})
we need one more step of integrating by parts. This step produces an
extra contribution to the term proportional to $dx/y$ as observed in the 
explicit computations in the text.

\vskip10mm
\noindent
{\Large\bf Appendix B}

\renewcommand{\theequation}{B.\arabic{equation}}\setcounter{equation}{0}

\vskip3mm\noindent
In this appendix, we present the explicit form of characteristic polynomials
$P_G^\rep(t)$ 
for $D_4$, $E_6$, $E_7$ and $E_8$ from which one can read off the relation
between the Casimir invariants and the deformation parameters. 

First of all, the characteristic polynomial for ${\bf 28}$ (adjoint) of $D_4$
reads
\begin{eqnarray}
P_{D_4}^{\bf 28}(t)  
&=& 
t^4 \biggl(
t^{24} - 18 w_2 t^{22} + 135 w_2^2 t^{20}
+ (12 \tilde{w}_4 w_2 - 24 w_6 - 552 w_2^3) t^{18}  \cr
& &
+(1359 w_2^4-10 \tilde{w}_4^2-114w_2^2 \tilde{w}_4+30w_4^2+198w_2 w_6) t^{16}
+\cdots\cdots \biggr) .
\end{eqnarray}
Next, we give the characteristic polynomial for 
${\bf 27}$ of $E_6$:
\begin{eqnarray}
P_{E_6}^{\bf 27}(t)
&=&
t^{27} + 12 w_2 t^{25}+60 w_2^2 t^{23} - 48 w_5 t^{22} +
(168 w_2^3 + 96 w_6) t^{21}\cr
& & 
 - 336 w_5 w_2 t^{20} +
(294 w_2^4 + 528 w_2 w_6 + 480 w_8) t^{19} \cr
& &
 - (1008 w_2^2 w_5  + 1344 w_9) t^{18}\cr
& &
+ (336 w_2^5 + 1152 w_2^2 w_6 + 2304 w_2 w_8 + 144 w_5^2) t^{17}\cr
& &
- (1680 w_2^3 w_5 + 5568 w_2 w_9 + 768 w_5 w_6) t^{16} \cr
& & 
+ (252 w_2^6 + 1200 w_2^3 w_6 + 4768 w_2^2 w_8 + 608 w_2 w_5^2\cr
& &
- 1248 w_6^2 + 17280 w_{12}) t^{15}
 + \cdots\cdots\cdots,
\end{eqnarray}
while $P_{E_6}^{\bf \overline{27}}(t)$ is obtained by letting
$w_5 \rightarrow -w_5$ and $w_9 \rightarrow -w_9$.

For ${\bf 78}$ (adjoint) of $E_6$ we have
\begin{eqnarray}
P_{E_6}^{\bf 78}(t)
&=&
t^6 \biggl(
t^{72}+48 w_2 t^{70}+1080 w_2^2 t^{68} + (15152 w_2^3-576 w_6) t^{66}\cr
& &
+(8640 w_8+148764 w_2^4-22752 w_6 w_2) t^{64} \cr
& &
+(297216 w_8 w_2-418176 w_2^2 w_6+1087632 w_2^5+6048 w_5^2) t^{62}\cr 
& &
+(-1071360 w_{12}+4749888 w_2^2 w_8+55872 w_6^2-4760352 w_2^3 w_6\cr
& &
+187584 w_5^2 w_2+6152776 w_2^6) t^{60} + \cdots\cdots\cdots \biggr) .
\end{eqnarray}

The characteristic polynomials for ${\bf 56}$ and ${\bf 133}$ (adjoint) of 
$E_7$ are given by 
\begin{eqnarray}
P_{E_7}^{\bf 56}(t)
&=&
t^{56} - 2^2\cdot 36 w_2   t^{54} + 2^4\cdot 594 w_2^2   t^{52}
+ 2^6  (72 w_6 - 6084 w_2^3) t^{50}\cr
& & 
+ 2^8 (-1800 w_2 w_6 + 60 w_8 + 43875 w_2^4) t^{48}\cr
& &
+ 2^{10} (21600 w_2^2 w_6 - 504 w_{10} - 1008 w_8 w_2 - 238680 w_2^5) t^{46}\cr
& &
+ 2^{12} (-540 w_{12} + 1022580 w_2^6 + 7008 w_2^2 w_8 + 10344 w_2 w_{10} \cr
& &
   -165600 w_6 w_2^3 + 540 w_6^2) t^{44} \cr
& &
+ 2^{14} (910800 w_2^4 w_6 - 3552120 w_2^7 + 7944 w_{12} w_2 - 1092 w_8 w_6 \cr
& &
-100824 w_2^2 w_{10}
- 20592 w_8 w_2^3 + 3828 w_{14} - 11592 w_2 w_6^2) t^{42}\cr
& &
+ 2^{16} (-49284 w_2^2 w_{12} + 630 w_8^2 + 620424 w_{10} w_2^3 
          + 22716 w_2 w_6 w_8\cr
& &
   -3825360 w_6 w_2^5 - 63468 w_2 w_{14} + 10212345 w_2^8 - 3528 w_{10} w_6\cr
& &
   -38808 w_2^4 w_8 + 118692 w_2^2 w_6^2) t^{40} \cr
& &
+ 2^{18} (683760 w_2^5 w_8 - 12656 w_8^2 w_2 - 24667500 w_2^9 + 1848 w_6^3\cr
& &
  -771120 w_2^3 w_6^2 - 29496 w_{18} + 489288 w_2^2 w_{14} 
- 2702280 w_2^4 w_{10}
\cr
& &
  +8760 w_{12} w_6 - 224040 w_2^2 w_6 w_8 + 5024 w_{10} w_8 
  +12751200 w_2^6 w_6\cr
& &
  +61824 w_2 w_6 w_{10}+145200 w_{12} w_2^3) t^{38} + \cdots\cdots\cdots,
\end{eqnarray}
\begin{eqnarray}
P_{E_7}^{\bf 133}(t)
&=&
t^7 \biggl(
t^{126}-2^2 \cdot 108 w_2 t^{124} + 2^4 \cdot 5616 w_2^2 t^{122} 
+ 2^6 (-144 w_6-187200 w_2^3) t^{120}\cr
& &
+ 2^8 (14400 w_2 w_6+600 w_8+4492800 w_2^4) t^{118} \cr
& &
+ 2^{10} 
(-691200 w_2^2 w_6+1008 w_{10}-54144 w_8 w_2-82667520 w_2^5) t^{116}\cr
& & 
+ 2^{12} (16200 w_{12}+1212456960 w_2^6+2337792 w_2^2 w_8-78144 w_2 w_{10}\cr
& &
+ 21196800 w_6 w_2^3+5400 w_6^2) t^{114}\cr
& &
+2^{14} 
(-466329600 w_2^4 w_6-14549483520 w_2^7-1345728 w_{12} w_2-59736 w_8 w_6\cr
& &
+ 2809344 w_2^2 w_{10}-64272384 w_8 w_2^3+71544 w_{14}
 -518976 w_2 w_6^2) t^{112}\cr 
& &
+ 2^{16} (7834337280 w_6 w_2^5+145494835200 w_2^8+53671104 w_2^2 w_{12}\cr
& &
+ 4816944 w_2 w_6 w_8-61360128 w_{10} w_2^3+1263144960 w_2^4 w_8\cr
& &
  -5463792 w_2 w_{14}+23770368 w_2^2 w_6^2+27900 w_8^2
  -210672 w_{10} w_6) t^{110}\cr
& &
+ 2^{18} (2679792 w_{18}+199042752 w_{14} w_2^2-331440 w_{12} w_6
-1368980736 w_{12} w_2^3\cr
& &
-339328 w_{10} w_8+14852352 w_{10} w_6 w_2
 +886013952 w_{10} w_2^4-1824128 w_8^2 w_2\cr
& &
-184786752 w_8 w_6 w_2^2
-18885672960 w_8 w_2^5+252624 w_6^3-690619392 w_6^2 w_2^3\cr
& &
 -104457830400 w_6 w_2^6-1228623052800 w_2^9) t^{108} 
+ \cdots\cdots\cdots \biggr) .
\end{eqnarray}

Finally, we write the  characteristic polynomial 
\begin{equation}
P_{E_8}^{\bf 248}(t) = t^8 \; \sum_{n=0}^{240} c_n t^n 
\end{equation}
for ${\bf 248}$ (adjoint) of $E_8$.
In this case, we show only eight coefficients which
are  sufficient to determine the relation between the Casimirs
and the deformation parameters. They are given by
\begin{eqnarray}
c_{238} 
&=&
2^2 \cdot 60 w_2,\cr
c_{232} 
&=&
2^8 (478170 w_2^{4} + 720 w_8),\cr
c_{228} 
&=&
2^{12} (47747700 w_2^{6} + 15120 w_{12} + 1030320 w_2^{2} w_8),\cr
c_{226} 
&=&
2^{14} (361791144 w_2^{7} + 79200 w_{14}
 + 17858880 w_2^{3} w_8 + 753840 w_2 w_{12}),\cr
c_{222} 
&=&
2^{18} (13257944700 w_2^{9} + 2620800 w_{18} + 293378400 w_{2}^{3} w_{12} 
+ 5240640 w_2 w_8^{2}\cr
& &
+ 2277007200 w_2^{5} w_8 + 96593280 w_2^{2} w_{14}),\cr
c_{220} 
&=& 
2^{20} (11040480 w_{20} + 65910925080 w_2^{10} + 123173712 w_2 w_{18} 
+ 1545977808 w_2^{3} w_{14} \cr
& & 
+ 3431681424 w_2^{4} w_{12} + 18595558800 w_2^{6} w_8 
+ 128513424 w_2^{2} w_8^{2} \cr
& &
+ 2492208 w_8 w_{12}),\cr
c_{216}
&=&
2^{24} (419237280 w_{24} + 1153992168420 w_2^{12} - 35394408 w_{12}^{2} 
+ 4551984 w_8^{3} \cr
& &
+ 11556147624 w_2^{2} w_{20} + 42618310896 w_2^{3} w_{18} 
+ 168171466680 w_2^{5} w_{14} \cr
& &
+ 234127252800 w_2^{6} w_{12} + 24236204440 w_2^{4} w_8^{2} 
+ 2516521104 w_2^{2} w_8 w_{12} \cr
& &
+ 749135368800 w_2^{8} w_8 + 387688872 w_{14} w_8 w_2),\cr
c_{210} 
&=&
2^{30} (65945880000 w_{30} + 39472177353840 w_2^{15} 
+ 5508702912024 w_{24} w_2^{3} \cr
& &
+ 15986969259936 w{20} w_2^{5} - 3209804640 w_{20} w_8 w_2 \cr
& &
+ 28604105079744 w_{18} w_2^{6} 
+ 234901945584 w_{18} w_8 w_2^{2} \cr
& &
- 4971002400 w_{18} w_{12} - 18339605640 w_{14}^{2} w_2
+ 250521815304 w_{14} w_{12} w_2^{2} \cr
& &
- 422863200 w_{14} w_{8}^{2} + 1528645019808 w_{14} w_8 w_2^{4} \cr
& &
+ 43713099157440 w_{14} w_2^{8} - 521644115232 w_{12}^{2} w_2^{3} \cr
& &
+ 8050693680 w_{12} w_8^{2} w_2 + 3139744251456 w_{12} w_8 w_2^{5} \cr
& &
+ 36016821822240 w_{12} w_2^{9} + 71061462976 w_8^{3} w_2^{3} \cr
& &
+ 10597571701120 w_8^{2} w_2^{7} + 68920453929600 w_8 w_2^{11}).
\end{eqnarray}

\newpage



\begin{thebibliography}{99}

\bibitem{Sen1} A. Sen, Nucl. Phys. {\bf B475} (1996) 562, hep-th/9605150

\bibitem{BDS} T. Banks, M. Douglas and N. Seiberg, 
Phys. Lett. {\bf B387} (1996) 278, hep-th/9605199

\bibitem{Sei} N. Seiberg, Phys. Lett. {\bf B388} (1996) 753, hep-th/9608111

\bibitem{MN1} J.A. Minahan and D. Nemeschansky,
Nucl. Phys. {\bf B482} (1996) 142, hep-th/9608047

\bibitem{MN2} J.A. Minahan and D. Nemeschansky,
Nucl. Phys. {\bf B489} (1997) 24, hep-th/9610076

\bibitem{SW2} N. Seiberg and E. Witten, Nucl. Phys. {\bf B431} (1994) 484, 
hep-th/9408099

\bibitem{AD} P.C. Argyres and M.R. Douglas, Nucl. Phys. {\bf B448} (1995) 93,
hep-th/9505062

\bibitem{APSW} P.C. Argyres, R.N. Plesser, N. Seiberg and E. Witten,
Nucl. Phys. {\bf B461} (1996) 71, hep-th/9511154

\bibitem{EHIY} T. Eguchi, K. Hori, K. Ito and S.-K. Yang, Nucl. Phys.
{\bf B471} (1996) 430, hep-th/9603002

\bibitem{MNS} A. Mikhailov, N. Nekrasov and S. Sethi,
Nucl. Phys. {\bf B531} (1998) 345, hep-th/9803142

\bibitem{dWHIZ} O. DeWolfe, T. Hauer, A. Iqbal and B. Zwiebach,
Nucl. Phys. {\bf B534} (1998) 261, hep-th/9805220

\bibitem{Vafa} C. Vafa, Nucl. Phys. {\bf B469} (1996) 403, hep-th/9602022

\bibitem{Kod} K. Kodaira, Ann. Math. {\bf 77} (1963) 563; 
Ann. Math. {\bf 78} (1963) 1

\bibitem{DM} K. Dasgupta and S. Mukhi, Phys. Lett. {\bf B423} (1998) 261, 
hep-th/9711094

\bibitem{Joh} A. Johansen, Phys. Lett. {\bf B395} (1997) 36, hep-th/9608186

\bibitem{GZ} M.R. Gaberdiel and B. Zwiebach, Nucl. Phys. {\bf B518} (1998) 151,
hep-th/9709013

\bibitem{DHIZ} O. DeWolfe, T. Hauer, A. Iqbal and B. Zwiebach,
{\it Uncovering the Symmetries on {\rm [$p,q$]} 7-branes: Beyond the Kodaira
Classification}, hep-th/9812028

\bibitem{dWZ} O. DeWolfe and B. Zwiebach, Nucl. Phys. {\bf B541} (1999) 509,
hep-th/9804210

\bibitem{Sen2} A. Sen, Phys. Rev. {\bf D55} (1997) 2501, hep-th/9608005

\bibitem{SW1} N. Seiberg and E. Witten, Nucl. Phys. {\bf B426} (1994) 19, 
hep-th/9407087

\bibitem{Ferr} F. Ferrari, Phys. Rev. Lett. {\bf 78} (1997) 795,
hep-th/9609101

\bibitem{BrSt} A. Brandhuber and S. Stieberger, Nucl. Phys. {\bf B488} (1997)
199, hep-th/9610053

\bibitem{BF} A. Bilal and F. Ferrari, Nucl. Phys. {\bf B516} (1998) 175,
hep-th/9706145

\bibitem{MVW} E. Martinec, Phys. Lett. {\bf 217B} (1989) 431; \\
C. Vafa and N.P. Warner, Phys. Lett. {\bf 218B} (1989) 51; \\
See, for a review, N.P. Warner, {\it $N=2$ Supersymmetric Integrable
Models and Topological Field Theories}, hep-th/9301088

\bibitem{EYY} T. Eguchi, Y. Yamada and S.-K. Yang, 
Mod. Phys. Lett. {\bf A8} (1993) 1627, \\ hep-th/9302048

\bibitem{EY}
T. Eguchi and S.-K. Yang, Phys. Lett. {\bf B394} (1997) 315, hep-th/9612086

\bibitem{ItYa}
K. Ito and S.-K. Yang, Phys. Lett. {\bf B415} (1997) 45, hep-th/9708017

\bibitem{IY2}
K. Ito and S.-K. Yang, Int. J. Mod. Phys. {\bf A13} (1998) 5373, hep-th/9712018

\bibitem{Shio} T. Shioda, J. Math. Soc. Japan, {\bf 43} (1991) 673
and references therein. 

\bibitem{Ha} R. Hartshorne, {\it Algebraic Geometry}, Springer-Verlag (1977),
chapter V 4

\bibitem{TeYa4} S. Terashima and S.-K. Yang, 
Nucl. Phys. {\bf B537} (1998) 344, hep-th/9808022

\bibitem{TeYa3} S. Terashima and S.-K. Yang, Phys. Lett. {\bf B430} (1998) 102,
hep-th/9803014

\bibitem{ItYa1} K. Ito and S.-K. Yang, Phys. Lett. {\bf B366} (1996) 165,
hep-th/9507144

\bibitem{AGMZ} L. \'Alvarez-Gaum\'e, M. Mari\~no and F. Zamora, Int. J. Mod.
Phys. {\bf A13} (1998) 403, hep-th/9703072

\bibitem{MW} E.J. Martinec and N.P. Warner, Nucl. Phys. {\bf B459} (1996) 97,
hep-th/9509161

\bibitem{spect} V. Kanev, Proc. Symp. Math. {\bf 49} (1989) 627; \\
A.~McDaniel, Duke. Math. J. {\bf 56} (1988) 47

\bibitem{LW} W. Lerche and N.P. Warner, Phys. Lett. {\bf B423} (1998) 79,
hep-th/9608183

\bibitem{BeFa} O. Bergman and A. Fayyazuddin, 
Nucl. Phys. {\bf B531} (1998) 108, hep-th/9802033; 
Nucl. Phys. {\bf B535} (1998) 139, hep-th/9806011

\bibitem{Oh} Y. Ohtake, {\it String Junctions and the BPS Spectrum of N=2
SU(2) Theory with Massive Matters}, hep-th/9812227

\end{thebibliography}
\end{document}